\documentclass[a4paper,fleqn]{cas-dc}
\usepackage{amsmath}
\usepackage{amssymb}
\usepackage{siunitx}
\usepackage{savesym}
\usepackage{bm}
\savesymbol{fax}
\savesymbol{Hermaphrodite}
\usepackage{marvosym}

\usepackage[authoryear]{natbib}
\newcommand{\mSaturn}{\text{\Saturn}}
\begin{document}
\shorttitle{Hyperion rotation and orbit}
\shortauthors{Goldberg \& Batygin}
\title[mode = title]{Chaotic tides as a solution to the Hyperion problem}

\author[1]{Max Goldberg}[orcid=0000-0003-3868-3663]
\cormark[1]
\ead{mg@astro.caltech.edu}

\author[2]{Konstantin Batygin}[orcid=0000-0002-7094-7908]

\cortext[1]{Corresponding author}

\affiliation[1]{organization={Department of Astronomy, California Institute of Technology}, addressline={1200 E. California Blvd}, city=Pasadena, state=CA, postcode=91125, statesep={}, country=USA}

\affiliation[2]{organization={Division of Geological and Planetary Sciences, California Institute of Technology}, addressline={1200 E. California Blvd}, city=Pasadena, state=CA, postcode=91125, statesep={}, country=USA}

\begin{abstract}
The dynamics of the outer regular satellites of Saturn are driven primarily by the outward migration of Titan, but several independent constraints on Titan's migration are difficult to reconcile with the current resonant orbit of the small satellite Hyperion. We argue that Hyperion's rapid irregular tumbling greatly increases tidal dissipation with a steep dependence on orbital eccentricity. Resonant excitation from a migrating Titan is then balanced by damping in a feedback mechanism that maintains Hyperion's eccentricity without fine-tuning. The inferred tidal parameters of Hyperion are most consistent with rapid Titan migration enabled by a resonance lock with an internal mode of Saturn, but a scenario with only equilibrium dissipation in Saturn is also possible.
\end{abstract}


\maketitle

\section{Introduction}
Much like miniature planetary systems, the regular satellites of Saturn are expected to have originated on nearly coplanar and circular orbits within the circumplanetary disk or the planet's rings. Since their formation, tidal dissipation within Saturn has caused the moons to migrate outwards and encounter mean motion resonances with each other. In some cases, pairs of moons captured into these mean motion resonances and remain there today, while in others there is indirect evidence of the excitation caused by resonant encounters. As such, resonant dynamics offer a unique window into the system's evolutionary past. In the tightly packed inner saturnian system (i.e. interior to Titan), a complex web of resonances sets strict constraints on the relative migration of each moon \citep{Cuk2016,Cuk2023}. In contrast with the well-understood dynamical history of the inner moons, the Titan--Hyperion system is strikingly enigmatic and one of the most remarkable mysteries of Solar System dynamics. 

Hyperion, the only satellite in the large gap between Titan and Iapetus, is trapped in an exterior 4:3 mean-motion resonance with Titan. The origins of this orbital configuration have historically been attributed to an outwardly evolving Titan capturing Hyperion into commensurability \citep{Colombo1974}. This scenario is accompanied by specific consequences: preservation of the adiabatic invariant \citep{Henrard1982} implies that, assuming no dissipation within Hyperion, Titan must migrate $4\%$ in semi-major axis post-capture \citep{Cuk2013} and thus the tidal $Q$ of Saturn must be $Q_\mSaturn<1500$.\footnote{However, too much migration of Titan is problematic: if Saturn has especially strong dissipation ($Q_\mSaturn < 500$), Titan and Hyperion would have started wide of the 3:2 resonance and captured into the wrong resonance.}

There are, however, contradicting constraints on Titan's migration from Iapetus, the outermost regular satellite of Saturn. Iapetus lies just $0.4\%$ inside the 5:1 mean-motion resonance with Titan, implying a recent but significant dynamical interaction between the moons. During the 5:1 resonant encounter, the eccentricity and inclination of Iapetus evolve chaotically and Titan's migration must be rapid enough to avoid ejecting Iapetus \citep{Cuk2013,Polycarpe2018}. Evidently, the preservation of Iapetus necessitates rapid migration of Titan, while Hyperion's resonance demands slow, short-range migration.

Hyperion's rotational properties are equally remarkable, and constitute a unique example of stochastic rotation in the Solar System. \cite{Wisdom1984} predicted that it would be in a chaotically tumbling state before its rotation was directly observed. They argued that given its moderate eccentricity ($e\approx 0.1$) and highly elongated shape (seen in Voyager 2 images), regular rotation in the synchronous (1:1) or 3:2 spin-orbit state is impossible. Instead, a chaotic zone surrounds the 1:1 and 2:1 spin-orbit resonances and Hyperion's spin vector evolves over timescales of a few orbital periods. In addition, \cite{Wisdom1984} demonstrated that much of the parameter space is attitude unstable, so that an initial small obliquity is quickly amplified and rotation inevitably occurs on all three axes. Early ground-based light curve observations by \cite{Klavetter1989} confirmed non-periodic rotation and suggested rotation at roughly the synchronous rate.

Despite the remarkable predictive power of \cite{Wisdom1984}'s calculations, images and light curves taken during the Voyager 2 and Cassini visits to the saturnian system demonstrated that Hyperion was rotating much faster than expected: roughly 4.2 times the synchronous rate \citep{Thomas1995,Harbison2011}. Rotation mostly occurs around the longest axis and is quasi-regular. Nevertheless, the wobble and precession are indeed clearly chaotic, with typical Lyapunov times of several orbital periods, as measured by \citet{Black1995} and \citet{Harbison2011}. Using numerical simulations, \cite{Black1995} showed that this state was not an unexpected outcome---initialized near the synchronous state, Hyperion would irregularly alternate between slower chaotic tumbling and more rapid quasi-regular rotation, the latter being the state actually observed by \citet{Klavetter1989} and \citet{Harbison2011}.

In light of the discrepancies between the predicted rotational behavior of Hyperion and its observed state, as well as an unclear relationship between the Titan migration rate and the Titan-Hyperion mean-motion resonance, a complete understanding of the Hyperion problem remains elusive.
Previous work has generally considered the rotation to be solely a \textit{consequence} of the orbit and neglected the impact of dissipation within Hyperion on its orbital evolution. We argue that tidal dissipation within Hyperion is non-negligible due to its rapid rotation, and in fact mediates its orbital eccentricity growth despite resonant forcing from Titan's migration. As a result, several fine-tuning problems are avoided and the tidal quality factor of Hyperion can be estimated.
We begin in Section \ref{sec:rot} by studying the chaotic and quasi-regular rotation of Hyperion and calculating the resulting tidal dissipation. Then, in Section \ref{sec:orb}, we use this new picture of dissipation within Hyperion to set constraints on the range and rate of Titan's outward migration. Finally, Section \ref{sec:discussion} discusses the implications of this proposed dynamical history of Hyperion.

\section{Rotational dynamics of Hyperion}
\label{sec:rot}
\subsection{Numerical procedure}
To investigate its rotational dynamics, we numerically modeled the spin and orientation of Hyperion under the effect of Saturn's gravity. The satellite is assumed to be in a fixed elliptical orbit around Saturn with eccentricity $e$ and true anomaly $f$. The units are chosen such that the semi-major axis $a$ is unity, the orbital period is $2\pi$, and $G M_\mSaturn = 1$, where $M_\mSaturn$ is the mass of Saturn. Hyperion is modeled as an ellipsoid with principal moments of inertia $A<B<C$ and its spin is represented by $\omega_a$, $\omega_b$, and $\omega_c$, the projections of the spin vector on the principal axes, so that the total spin rate is $|\omega|=\sqrt{\omega_a^2+\omega_b^2+\omega_c^2}$. The spins evolve according to Euler's equations, 
\begin{align}
    \label{eq:euler}
    \dot{\omega}_a = \frac{B-C}{A}\left(\omega_b \omega_c -\frac{3}{r^3}\beta\gamma\right)\\
    \dot{\omega}_b = \frac{C-A}{B}\left(\omega_c \omega_a -\frac{3}{r^3}\gamma\alpha\right) \\
    \dot{\omega}_c = \frac{A-B}{C}\left(\omega_a \omega_b -\frac{3}{r^3}\alpha\beta\right)
\end{align}
in which the external torque is provided by the gradient of the gravitational field of Saturn \citep{Murray1999}. Here, $r$ is the instantaneous Hyperion--Saturn distance and $\alpha$, $\beta$, and $\gamma$ are the direction cosines between the principal axes and the direction of Saturn.

To represent the orientation of Hyperion, we use the quaternion formalism, which avoids the coordinate singularities that appear when using Euler angles \citep{Melnikov2020}. The four quaternion components $\lambda_0$, $\lambda_1$, $\lambda_2$, and $\lambda_3$ are normalized and evolve according to \citep{Arribas2006}
\begin{align}
\dot{\lambda}_0 &= \frac{1}{2}(-\lambda_1\omega_a - \lambda_2\omega_b - \lambda_3\omega_c)\\
\dot{\lambda}_1 &= \frac{1}{2}(\lambda_0\omega_a - \lambda_3\omega_b + \lambda_2\omega_c)\\
\dot{\lambda}_2 &= \frac{1}{2}(\lambda_3\omega_a + \lambda_0\omega_b - \lambda_1\omega_c)\\
\dot{\lambda}_3 &= \frac{1}{2}(-\lambda_2\omega_a + \lambda_1\omega_b + \lambda_0\omega_c).
\label{eq:quat}
\end{align}
The direction cosines are given by
\begin{align}
\alpha &= (\lambda_0^2 + \lambda_1^2 - \lambda_2^2 - \lambda_3^2)\cos f + 2(\lambda_0\lambda_3 + \lambda_1\lambda_2)\sin f\\
\beta &= 2(\lambda_1\lambda_2 - \lambda_0\lambda_3)\cos f + (\lambda_0^2 - \lambda_1^2 + \lambda_2^2 - \lambda_3^2) \sin f\\
\gamma &= 2(\lambda_0\lambda_2 + \lambda_1\lambda_3)\cos f + 2(-\lambda_0\lambda_1 + \lambda_2\lambda_3)\sin f.
\label{eq:cosines}
\end{align}
We use the moment of inertia parameters $A=0.314$, $B=0.474$, $C=0.542$ estimated by \cite{Harbison2011} from a Cassini shape model. We then numerically integrate Eqs.~\ref{eq:euler}--\ref{eq:quat} with a fifth-order Radau IIA method using relative and absolute tolerances of $10^{-6}$ and $10^{-10}$, respectively. We ran 10 integrations for $3\times 10^6$ orbits with orbital eccentricities ranging uniformly in log-space from $0.01$ to $0.631$. Each was started at synchronous rotation ($|\omega|=1$) but with an obliquity of $1^\circ$ to induce tumbling \citep{Black1995}. Although wobble damping \citep{Burns1973,Wisdom1987} may be relevant on such long timescales, the purpose of these simulations is to determine the range of typical rotational dynamics; long integrations are more likely to capture rare behavior and less likely to be trapped in ``small tributaries of the chaotic zone,'' as noticed by \citet{Wisdom1987}. For comparison, we also integrated the rotation of Hyperion from its observed state on 2005--06--10 for $10^3$ orbits, using the orientation and spin vectors reported by \cite{Harbison2011}. 

\subsection{Rotational evolution of Hyperion}
\begin{figure*}
    \centering
    \includegraphics[width=\textwidth]{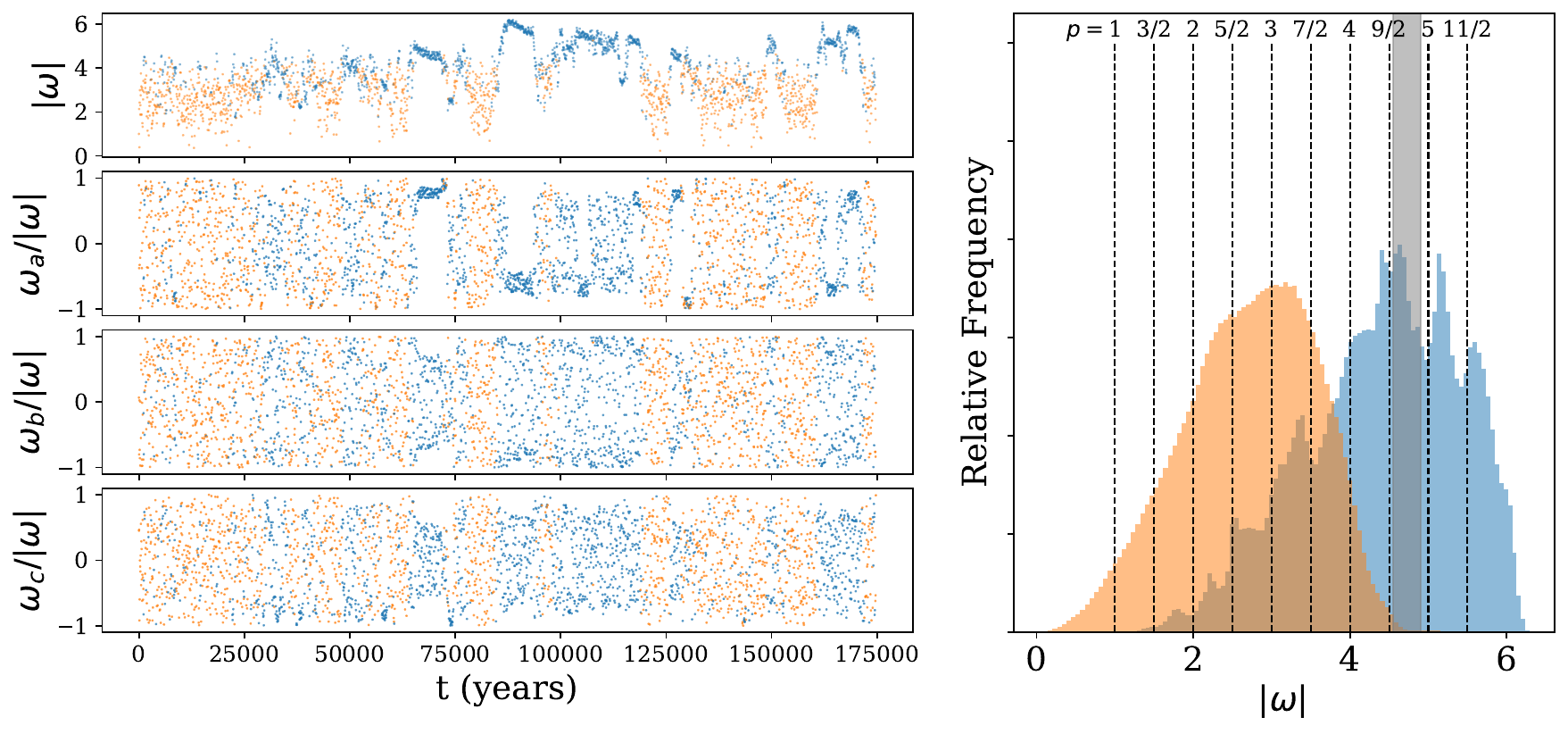}
    \caption{Left: A typical integration of the rotational equations of Hyperion for $3\times 10^6$ orbits, starting from a nearly synchronous state and a realistic orbital eccentricity of $0.1$. Chaotic tumbling (shaded orange) intermittently gives way to quasi-regular rotation (blue). Right: the distribution of $|\omega|$ in each state. The two distributions are shown to scale relative to each other. Dashed vertical lines mark spin-orbit resonances. The shaded gray region is the $2\sigma$ range of Hyperion's rotation speed in its observed state in 2005.}
    \label{fig:rot}
\end{figure*}

One example of the longer integrations is shown in Figure \ref{fig:rot}, where we have chosen $e=0.1$. Hyperion begins in a chaotic tumbling state, shaded in orange on the plot, but intermittently passes through quasi-regular states, shaded in blue. As noted by \cite{Black1995}, quasi-regular states are typically associated with rotation primarily on axes $a$ or $c$ ($b$ is not stable owing to the intermediate axis theorem). The right panel of Figure~\ref{fig:rot} shows the distribution of angular speeds in these two regimes. Chaotic tumbling is smoothly distributed across all values of $|\omega|\lesssim5$. However, quasi-regular rotation is faster and dominated by peaks at discrete values of $|\omega|$ that correspond to half-integer spin-orbit resonances. At higher $|\omega|$, the peaks lie slightly wide of exact resonance. In particular, the state of Hyperion in 2005 (shaded in gray) matches the highest peak, which appears to be associated with the $9/2$ resonance.

The clustering behavior near spin-orbit resonance is probably a consequence of the resonant ``sticking'' effect \citep{Karney1983,Meiss1992,Shevchenko1999}. In the vicinity of the separatrix that bounds large resonant islands, there are numerous small islands of secondary resonances. Chaotic trajectories which wander near the separatrix may be caught in one of these islands, which necessarily lie in proximity to the resonance. The trajectory will then evolve very slowly through action space and the rotation will be in a quasi-regular state for an extended duration.

Simulations at other eccentricities were qualitatively similar to the $e=0.1$ case. Alternation between chaotic tumbling and quasi-regular rotation near spin-orbit resonances was observed at all the eccentricities we tested. The tumbling state accounted for 30--60\% of the total duration, with no clear dependence of that fraction on eccentricity.

\begin{figure}
    \centering
    \includegraphics[width=0.5\textwidth]{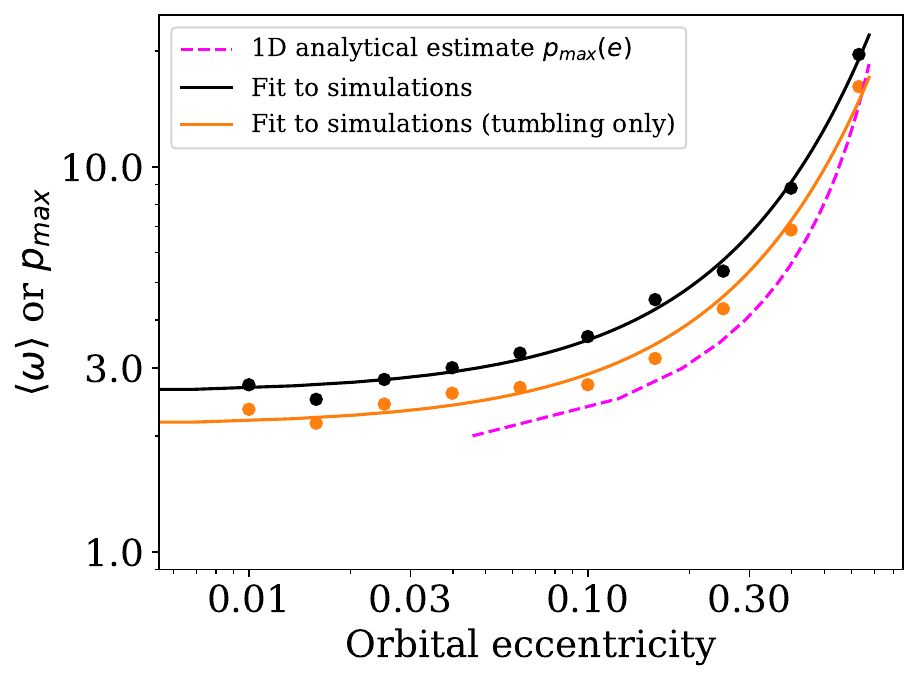}
    \caption{Average rotation speed of Hyperion as a function of its orbital eccentricity. Black dots are the full long integrations and the black line is an fit with an exponential dependence on eccentricity. Orange is the same but considering only the chaotic tumbling state, removing the quasi-regular rotation (see Figure~\ref{fig:rot}). The magenta curve is the analytical estimate of $p_{max}$ from solving Eq.~\ref{eq:res_overlap}. The analytic solution, despite being offset from the numerical results by a factor of $\sim 2$, captures the qualitative behavior of $\langle\omega\rangle$ as a function of $e$.}
    \label{fig:omega_vs_e}
\end{figure}

However, the typical rotation speed in the long integrations shows a strong dependence on orbital eccentricity \citep{Wisdom1987,Quillen2020}. Denoting the time average of $|\omega|$ as $\langle\omega\rangle$, the typical $\langle\omega\rangle$ was much higher for higher $e$ during both chaotic tumbling and quasi-regular rotation. Figure \ref{fig:omega_vs_e} shows the $\langle\omega\rangle$ as a function of eccentricity for the long simulations. We also computed the mean $\langle\omega\rangle$ during the chaotic tumbling only, by removing the times with quasi-regular motion as in Figure \ref{fig:rot}. Both are fit well by an exponential dependence on $e$. Without removing quasi-regular motion, we find $\langle\omega\rangle\approx2.59\times 1.42^{e/0.1}$. Considering chaotic tumbling only, we obtain $\langle\omega\rangle\approx 2.03\times 1.38^{e/0.1}$. Finally, for reasons that will become apparent in Section~\ref{sec:tides}, we fit the fourth root of the time average of $|\omega|^4$ as a function of $e$ in the same way, finding $\langle\omega^4\rangle^{1/4}\approx 2.88\times 1.37^{e/0.1}$ and $\langle\omega^4\rangle^{1/4}\approx 2.27\times 1.38^{e/0.1}$ for all rotation and chaotic tumbling, respectively.

\subsection{Analytical rotation model}
To qualitatively understand the eccentricity dependence of $\langle\omega\rangle$ seen in our simulations, it is instructive to consider a simplified one-dimensional model of Hyperion's spin, even though its rotation is fully three-dimensional. This simplification ignores the obliquity and spin precession of Hyperion, but can be studied analytically in much more detail. Following \cite{Wisdom1984}, assume the satellite's spin axis is perpendicular to its orbital plane. The orientation of the satellite in the inertial frame is given by $\theta$.  Then, $\theta$ evolves according to
\begin{equation}
    \frac{d^2 \theta}{dt^2} + \frac{\omega_0^2}{2r^3}\sin 2(\theta-f) = 0
    \label{eq:1d_rot}
\end{equation}
where $\omega_0^2=3(B-A)/C$ \citep{Goldreich1966}. Equation \ref{eq:1d_rot} is unwieldy because $r$ and $f$ are complicated functions of time. However, it can be expanded via a Fourier series into
\begin{equation}
    \frac{d^2 \theta}{dt^2} + \frac{\omega_0^2}{2}\sum_{p=-\infty}^\infty H(p,e) \sin (2\theta - 2pt) = 0
    \label{eq:1d_fourier}
\end{equation}
where $p$ is a half-integer and the $H(p,e)$ are coefficients given by
\begin{equation}
    H(p,e)=\frac{1}{2\pi}\int_0^{2\pi}\frac{1}{r^3} \cos(2pt-2f) dt,
\end{equation}
which, for $e\ll 1$ and $p\lesssim 5$ is of order $H(p,e)\sim 2^{2p-1}e^{2p-2}$ \citep{Dobrovolskis1995}. As is well-known \citep{Goldreich1966}, Equation \ref{eq:1d_fourier} points at the existence of a discrete set of spin-orbit resonances in which $d\theta/dt\approx p$. For example, the 3:2 spin-orbit state of Mercury corresponds to $p=3/2$, or three rotations (in the inertial frame) for every two orbits \citep{Goldreich1966}. The half-width of the spin-orbit resonance in frequency space is $\omega_0 \sqrt{H(p,e)}$, increasing with $e$ and decreasing with $p$. According to the resonance overlap criterion \citep{Chirikov1979}, chaotic behavior arises when neighboring resonances, (whose widths can be calculated to leading order as if they were unperturbed by each other), would overlap. Thus, chaos will appear around the $p$ and $p+1/2$ resonances if
\begin{equation}
\omega_0\sqrt{H(p,e)} + \omega_0\sqrt{H(p+1/2,e)} \gtrsim \frac{1}{2}.
\label{eq:res_overlap}
\end{equation}
\cite{Wisdom1984} use Eq.~\ref{eq:res_overlap} and the two widest resonances, $p=1$ and $p=3/2$, to generate a general condition for the existence of a broad chaotic region and argue that Hyperion must be in it.

Turning this argument around, we can also ask, for a given $e$ and $\omega_0$, what is the highest $p_\text{max}$ for which resonances overlap such that there is a chaotic sea surrounding the $p_\text{max}$ and $p_\text{max}+1/2$ resonances? Because of the dependence of $H(p,e)$ on $p$, the chaotic sea will also extend for at least $1\leq p \leq p_\text{max}+1/2$, and a trajectory initialized near $p=1$ will eventually explore up to $p_\text{max}$ ergodically. We solve Eq.~\ref{eq:res_overlap} numerically for $e$ by selecting a $p_\text{max}$ and taking $\omega_0=0.94$, corresponding to the values of $(A,B,C)$ we used in the 3D simulations. The result is shown as the magenta curve in Figure~\ref{fig:omega_vs_e}. The expression for $H(p,e)$ ensures that the size of the chaotic sea, and thus $p_\text{max}$, grow steeply with $e$ \citep{Wisdom1987}. Our analytical model closely matches the steep dependence on $e$ found in the numerical simulations, although the 1D model consistently underestimates $\langle\omega\rangle$. Evidently, the chaotic region is larger in 3D, and thus emerges at a smaller eccentricity for a given $p$. Indeed, the notion that the onset of chaos occurs earlier in systems with more degrees of freedom is qualitatively expected \citep[see, e.g.][]{Morbidelli2002}.

The steep dependence of Hyperion's rotation on its orbital eccentricity has an important consequence. Hyperion's eccentricity is resonantly excited by an outwardly migrating Titan, and in the absence of additional forces, increases monotonically. Hence, we expect that the spin rate of Hyperion has \textit{grown} over time, in contrast to most bodies in the Solar System.

\subsection{Tidal dissipation}
\label{sec:tides}
Some of the energy of the time-varying tidal torque is dissipated within Hyperion. The two main effects of the dissipation are despinning and eccentricity damping. The despinning timescale at the present orbit is roughly of the order the age of the Solar System \citep{Wisdom1984}. The eccentricity damping timescale is usually much larger than the despinning timescale and has therefore been ignored for Hyperion. We will examine this in more detail.

For a synchronously rotating satellite with low eccentricity, the eccentricity damping rate is given by
\begin{equation}
\tau_{e,\text{sync}}^{-1}\equiv -\frac{1}{e}\frac{de}{dt}=\frac{21}{2}\frac{k_{2,\text{H}}}{Q_\text{H}}\frac{M_\mSaturn}{M_\text{H}}\left(\frac{R_\text{H}}{a_\text{H}}\right)^5n_\text{H}
\label{eq:tau_esync}
\end{equation}
where $k_{2,\text{H}}$ is the tidal Love number of Hyperion, $Q_\text{H}$ is its tidal quality factor, $R_\text{H}$ is the average radius of Hyperion, and $M_\text{H}$, $a_\text{H}$ and $n_\text{H}$ are the mass, semi-major axis and mean motion of Hyperion, respectively \citep{Goldreich1966a}. While the exact values of $k_{2,\text{H}}$ and $Q_\text{H}$ are unknown, Hyperion is believed to be a rubble pile with high internal porosity \citep{Thomas2007}. \cite{Goldreich2009} suggest that for such an object, $k_{2,\text{H}}\lesssim 1\times 10^{-3}$ and $Q_\text{H}\lesssim 100$. Assuming $k_{2,\text{H}}$ is at the upper bound, $R_\text{H}=150$ km, $M_\mSaturn/M_\text{H}=1.0\times 10^8$, and Hyperion's current period of $21.28$ d, we obtain 
\begin{equation}
\tau_{e,\text{sync}}\approx 8 \times 10^{13} \left(\frac{Q_\text{H}}{100}\right) \text{yr},
\end{equation}
much longer than the age of the Solar System.

However, Hyperion is manifestly \textit{not} rotating synchro-nously. In non-synchronous rotation, the entirety of the tide is raised and lowered during one cycle, greatly enhancing the dissipation of energy. \cite{Burns1973} argue that the energy dissipated in a non-synchronous rotator per orbit is, to an order of magnitude,
\begin{equation}
\Delta E \sim \frac{|\omega|^4 R_\text{H}^5}{G}\frac{k_{2,\text{H}}}{Q_\text{H}}
\end{equation}
where we have written the equation in terms of $k_{2,\text{H}}$ rather than the rigidity \citep{Goldreich2009}. Energy dissipated during chaotic tumbling should be at least comparable to, if not much larger than this estimate \citep{Wisdom1987,Brasser2020}. Because angular momentum is conserved, this dissipation must drive circularization of the orbit, and thus the eccentricity damping rate for Hyperion's irregular rotation, $\tau_{e,\text{H}}^{-1}$ will be enhanced over the synchronous rate by roughly
\begin{equation}
\frac{\tau_{e,\text{H}}^{-1}}{\tau_{e,\text{sync}}^{-1}} \sim \frac{1}{e_\text{H}^2}\frac{\langle\omega^4\rangle}{n_\text{H}^4}.
\label{eq:damp_enhance}
\end{equation}

For current values of Hyperion, $e\approx0.1$ and $\langle\omega^4\rangle^{1/4}/n_\text{H}\approx4$ and the enhancement is $\sim 2\times 10^{4}$. \citet{Quillen2020} performed simulations of a viscoelastic model of the Martian satellites Phobos and Deimos and confirmed that the energy dissipation rate during episodes of rapid tumbling was larger than the dissipation during synchronous rotation by 3 to 5 orders of magnitude. In their case, the process is naturally quenched as the eccentricity is damped, rotation slows, and the satellites capture into a synchronous state. In contrast, Hyperion's eccentricity is continuously excited by a resonant interaction with Titan (Section \ref{sec:orb}) and rapid rotation does not cease.

With the enhancement from rapid rotation, the expected eccentricity damping timescale of Hyperion is now $\tau_{e,\text{H}}\sim 4\times10^{9}$ yr, of order the age of the Solar System. While this estimate should not be taken to be exact \citep[see][]{Wisdom1987} because of order-unity constants dropped by \cite{Burns1973}, the result is that circularization of Hyperion's orbit can no longer be ignored despite its considerable distance from Saturn. In addition, because of the exponential dependence of $|\omega|$ on $e$, the damping timescale varies considerably with $e$, in contrast to the synchronous timescale, which is independent of $e$. This is the critical piece coupling Hyperion's rotation to its orbital history.

\section{Orbital dynamics of the Titan--Hyperion system}
\label{sec:orb}
The presence of the 4:3 mean-motion resonance between Hyperion and Titan is usually interpreted as resulting from the outward migration of Titan \citep{Colombo1974,Cuk2013}.  The resonance has a libration amplitude of $36^\circ$, with a forced eccentricity of $0.1$, and one possible explanation for this state is the expansion of Titan's orbit by $4\%$ since the initial encounter with the resonance \citep{Cuk2013}. 

In light of the orbital coupling discussed above, however, it is critical to exmaine alternative scenarios. Much in the same manner as the Moon recedes from the Earth due to tides raised on the ocean, Titan migrates outward because it raises a tidal bulge on Saturn. As Saturn rotates faster than Titan orbits, the tidal bulge transfers angular momentum from Saturn's rotation to Titan's orbit. The rate of expansion of Titan's orbit is given by
\begin{equation}
\tau_{a,\text{Ti}}^{-1}= -\frac{1}{a_\text{Ti}}\frac{da_\text{Ti}}{dt}= -\frac{3k_{2,\mSaturn}}{Q_\mSaturn(n_\text{Ti})}\frac{M_\text{Ti}}{M_\mSaturn}\left(\frac{R_\mSaturn}{a}\right)^5n_\text{Ti}
\label{eq:tit_tid}
\end{equation}
where $Q_\mSaturn(n)$ is the tidal quality factor of Saturn at forcing frequency $n$ and $k_{2,\mSaturn}$ is the Love number of Saturn, which we take to be $0.382$ \citep{Lainey2020,Jacobson2022}. The sign convention, consistent with Equation \ref{eq:tau_esync}, means that $\tau_{a,\text{Ti}}<0$ corresponds to outward migration. Quantitative predictions for $Q_\mSaturn$ are challenging and highly dependent on Saturn's internal structure \citep{Ogilvie2004}. Observationally, $Q_\mSaturn(n)$ can be measured by observations of outward migration of the inner saturnian moons \citep{Lainey2012}. Interestingly, recent works have shown that $Q_\mSaturn$ is not the same for each of Saturn's moons, with Rhea especially having higher tidal dissipation \citep{Lainey2017}. Indeed, some tidal theories predict that $Q_\mSaturn$ should depend on the forcing frequency $n$, in some cases quite sensitively \citep{Ogilvie2004,Fuller2016,Terquem2021}. Therefore, the tidal quality factor relevant for Titan's migration cannot be assumed to be the same as the one measured for another of Saturn's moons. 

Recently, two groups have reported conflicting measurements of outward migration of Titan. \cite{Lainey2020} used astrometry and Cassini radio tracking to obtain $Q_\mSaturn(n_\text{Ti})=124^{+26}_{-19}$ ($3\sigma$ uncertainties) corresponding to a migration timescale of $\tau_{a,\text{Ti}}\approx \qty{10}{Gyr}$. They interpret this result as consistent with the resonant locking model, in which satellites couple to inertial modes within Saturn and migrate outwards as the interior of Saturn evolves over the lifetime of the Solar System. Long-range migration enabled by the resonant locking mechanism is also consistent with the expectation that Titan formed at or migrated to the inner edge of the circumplanetary disk, near a period of $\sim \qty{3}{d}$ \citep{Batygin2023b}. However, such rapid migration is disputed by \cite{Jacobson2022}, who uses a large corpus of tracking, astrometric, and other data to obtain $Q_\mSaturn(n_\text{Ti})=1224\pm 119$ ($1\sigma$ uncertainties), or $\tau_{a,\text{Ti}}\approx \qty{100}{Gyr}$, an order of magnitude slower than \cite{Lainey2020}.

The capture and evolution of Hyperion in its mean-motion resonance with Titan depends on the specifics of Titan's outward migration. For completeness, we consider two cases below: one in which Titan's migration is consistent with the results of \cite{Lainey2020}, and one in which it is consistent with the results of \cite{Jacobson2022}. 
\subsection{Analytical Results}
To leading order, the mean motion and eccentricity of Titan and Hyperion near the $4$:$3$ mean-motion resonance evolve according to \citep[e.g.,][]{Terquem2019}
\begin{align}
\dot{n}_\text{Ti} &= \frac{3n_\text{Ti}}{2\tau_{a,\text{Ti}}} + \frac{3n_\text{Ti}e_\text{Ti}^2}{\tau_{e,\text{Ti}}} \\
\dot{n}_\text{H} &= 12 n_\text{H}^2 \frac{M_\text{Ti}}{M_\mSaturn}(e_\text{Ti}f_1\sin{\phi_1} + e_\text{H} f'_2 \sin{\phi_2}) + \frac{3n_\text{H} e_\text{H}^2}{\tau_{e,\text{H}}}\\
\dot{e}_\text{Ti} &= -\frac{e_\text{Ti}}{\tau_{e,\text{Ti}}}\\
\dot{e}_\text{H} &= -n_\text{H} \frac{M_\text{Ti}}{M_\mSaturn} f'_2 \sin{\phi_2} - \frac{e_\text{H}}{\tau_{e,\text{H}}}
\end{align}
where we have assumed that Hyperion's mass and tidal migration rate are negligible. Here, $\phi_1=4\lambda_\text{H}-3\lambda_\text{Ti}-\varpi_\text{Ti}$ and $\phi_2=4\lambda_\text{H}-3\lambda_\text{Ti}-\varpi_\text{H}$ are the critical resonant angles and $f_1$ and $f'_2$ are order unity constants. To reduce these equations further, we note that the capture into resonance of Hyperion implies that $\dot{n}_\text{Ti}/n_\text{Ti}=\dot{n}_\text{H}/n_\text{H}$. Additionally, because Titan's pericenter precession is dominated by Saturn's $J_2$, $\phi_1$ circulates and the time average of $\sin\phi_1$ is zero. We thus obtain
\begin{equation}
    4 \dot{e}_\text{H} e_\text{H} = -3e_\text{H}^2 \tau_{e,\text{H}}^{-1} - \left(\frac{1}{2}\tau_{a,\text{Ti}}^{-1} + e_\text{Ti}^2\tau_{e,\text{Ti}}^{-1}\right)
\end{equation}
which, despite being simpler, must still be integrated numerically because of the distance and eccentricity dependence of the migration and damping timescales. Nevertheless, the competing effects of eccentricity excitation and damping suggest that we can compute an equilibrium eccentricity that Hyperion will tend towards. Setting $\dot{e}_\text{H}=0$, we find
\begin{equation}
e_\text{eq,H}^2 = -\frac13\left(\frac{1}{2}\tau_{a,\text{Ti}}^{-1} + e_\text{Ti}^2\tau_{e,\text{Ti}}^{-1}\right)\tau_{e,\text{H}}.
\end{equation}
Because $e_\text{Ti}\ll 1$ and for typical tidal processes, $|\tau_e|\sim |\tau_a|$, the second term in the brackets can be neglected. Recalling Eq.~\ref{eq:damp_enhance}, the equilibrium eccentricity can be translated into an equation for the equilibrium rotation rate of Hyperion,
\begin{equation}
\frac{\langle\omega^4\rangle}{n_\text{H}^4} \approx -\frac16\frac{\tau_{e,\text{sync}}}{\tau_{a,\text{Ti}}}.
\end{equation}

With the expected damping of Hyperion (Section \ref{sec:tides}) and the current rotation rate of Hyperion (averaged over the secular eccentricity cycle and the two rotation regimes), $\langle\omega^4\rangle^{1/4}/n_\text{H}\approx 4$, we find, assuming Hyperion is at its equilibrium eccentricity,
\begin{equation}
Q_\text{H} \approx 20 \left(\frac{|\tau_{a,\text{Ti}}|}{\qty[print-unity-mantissa = false]{1e10}{yr}}\right).
\end{equation}
Thus, if Hyperion is at its equilibrium eccentricity, the \cite{Lainey2020} Titan migration measurement implies $Q_\text{H}\sim 20$, while the \cite{Jacobson2022} value implies $Q_\text{H}\sim 200$. As we will see below, more accurate estimates of $Q_\text{H}$ are larger because Hyperion does not typically reach the equilibrium eccentricity.

\subsection{Numerical results}
As a means of testing our analytical theory, we ran a suite of N-body simulations modeling the outward migration of Titan and tidal dissipation of Hyperion resulting from its rapid chaotic tumbling. 
Previous work has coupled rotational and orbital integrations to study the spin-orbit evolution of irregular satellites \citep{Cuk2016a,Quillen2017,Quillen2020,Agrusa2021,Quillen2022}. However, our objective is to demonstrate the feasibility of resonant capture under enhanced tidal damping.
Accordingly, we do not repeat the rotational simulations but instead model tidal dissipation with an eccentricity damping term estimated using the results of Section \ref{sec:rot}.
Our numerical integrations use the \texttt{whfast} symplectic integrator in the \texttt{rebound} N-body package \citep{Rein2015}. The integrator timestep was chosen to be $1/20$ the initial orbital period of Titan. Additional forces for migration and eccentricity damping were incorporated with \texttt{reboundx} \citep{Tamayo2020a}. The integration includes the Sun and the $J_2$ moment of Saturn, to which is added the averaged orbits of the satellites interior to Titan. At each timestep, we compute the eccentricity damping of Hyperion using its instantaneous eccentricity, $e_\text{H}$. To accomplish this, we estimate the average rotation speed, $\langle\omega^4\rangle^{1/4}\approx 2.88\times 1.37^{e_\text{H}/0.1}$ (Section \ref{sec:rot}), and then use Eqs. \ref{eq:tau_esync} and \ref{eq:damp_enhance} to determine the enhanced eccentricity damping timescale, $\tau_{e,\text{H}}$. We also compute the migration rate and eccentricity damping rate of Titan at each timestep according to the prescriptions specified in the following sections. Then, during each ``kick'' step of the \texttt{whfast} algorithm, we apply an additional force to each satellite of
\begin{align}
    \bm{a}_{\text{damp,Ti}} &= -2\frac{(\bm{v}_\text{Ti} \cdot \bm{r}_\text{Ti})r_\text{Ti}}{(\bm{r}_\text{Ti}\cdot \bm{r}_\text{Ti}) \tau_{e,\text{Ti}}} -\frac{\bm{v}_\text{Ti}}{2\tau_{a,\text{Ti}}}  \\
    \bm{a}_{\text{damp,H}} &= -2\frac{(\bm{v}_\text{H} \cdot \bm{r}_\text{H})r_\text{H}}{(\bm{r}_\text{H}\cdot \bm{r}_\text{H}) \tau_{e,\text{H}}}
\end{align}
where $\bm{r}_i$ and $\bm{v}_i$ are the radius and velocity vector of the particle relative to Saturn, respectively \citep{Papaloizou2000}.

The simulation initial conditions were chosen to be compatible with available constraints. Although the age of Hyperion is not known precisely, its low orbital inclination implies that it, or its parent object, formed in the circumplanetary disk. High crater densities are also consistent with the notion that Hyperion is quite old \citep{Plescia1983,Bottke2023}. We thus ran the simulations over a timespan of $\qty{4.5}{Gyr}$. Hyperion was placed exterior to Titan with an initial period ratio of 1.35 in order to avoid capture into the wrong resonance. The initial eccentricity was varied between 0 and 0.05 and the initial inclination was set to 0 relative to Saturn's equator. The other orbital angles were randomized uniformly. 

It is important to note that the measurements of \cite{Lainey2020} and \cite{Jacobson2022} are only of Titan's current migration rate (baselines of $\approx \qty{150}{yr}$) and are not necessarily representative of the previous behavior of Titan. Accordingly, we attempt to construct a reasonable migration history of Titan in each case and include that in the simulation as described below.  To ensure feasible computational times, we sped up integrations by dividing the migration timescale of Titan and the eccentricity damping timescales of Titan and Hyperion by a common factor of $10^4$. We do not expect this to impact our results because the accelerated migration and damping timescales still greatly exceed the other dynamical timescales in this problem (which are $\lesssim \qty{100}{yr}$), ensuring that adiabaticity during the resonant encounter is preserved.

\subsubsection{Rapid Titan Migration}
\label{sec:fast}
First, we assume to be true the results of \cite{Lainey2020}, who find $\tau_{a,\text{Ti}}\approx -\qty{10}{Gyr}$ and argue that Titan is in a `resonant lock' with an internal mode of Saturn \citep{Fuller2016}. In such a regime, the migration of Titan is set by the interior evolution of Saturn, which unfolds roughly on the timescale of its age. Following \cite{Lainey2020}, we hypothesize
\begin{equation} 
\tau_{a,\text{Ti}}^{-1} \approx -\frac{B}{t}
\label{eq:tau_a_titan_res}
\end{equation}
where, to match the current $\tau_{a,Ti}$ measurement, $B\sim 1/3$. This equation has the solution
\begin{equation}
a_\text{lock}(t) = a_0 \left( \frac{t}{t_0} \right)^B \label{eq:a_lock}
\end{equation}
where $a_0$ is Titan's current semi-major axis and $t_0$ is Saturn's age. Of course, Eq.~\ref{eq:a_lock} cannot be strictly true, because $a_\text{lock}(0)=0$. Instead, a likely scenario is that Titan formed at an initial semi-major axis $a_i$ and remained there until some time $t_\text{lock}$, upon which point it caught into the resonant lock and Eq.~\ref{eq:a_lock} applies. Although $t_\text{lock}$ (equivalently $a_i$) is unknown, we find that our results do not depend significantly on its value.

As initial conditions, we choose $t_\text{lock}=1, 2$ or 3 Gyr, which correspond to an initial Titan semi-major axis of $12.28, 15.47$, and $17.71 R_\mSaturn$ respectively, and an initial Titan eccentricity of $0.04$. Titan migration occurs when $t>t_\text{lock}$, and we set the migration timescale to $\tau_{a,\text{Ti}}=-3t$ according to Eq.~\ref{eq:tau_a_titan_res}. The true eccentricity damping timescale of Titan is unknown, but is expected to be of the same order as the migration timescale \citep{Fuller2016}. Accordingly, we set $\tau_{e,\text{Ti}}=\tau_{a,\text{Ti}}$, so that Titan's final eccentricity is closed to its observed value of 0.029. 

\begin{figure}
    \centering
    \includegraphics[width=0.47\textwidth]{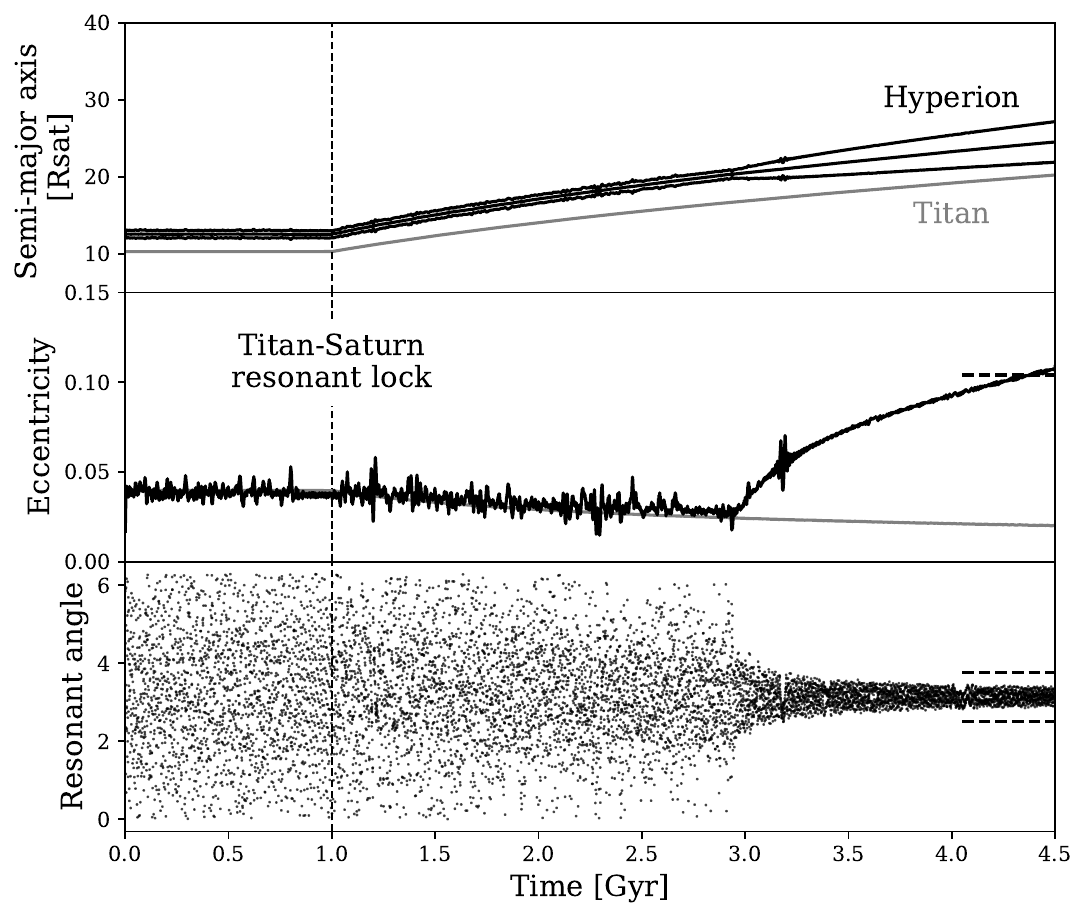}
    \includegraphics[width=0.47\textwidth]{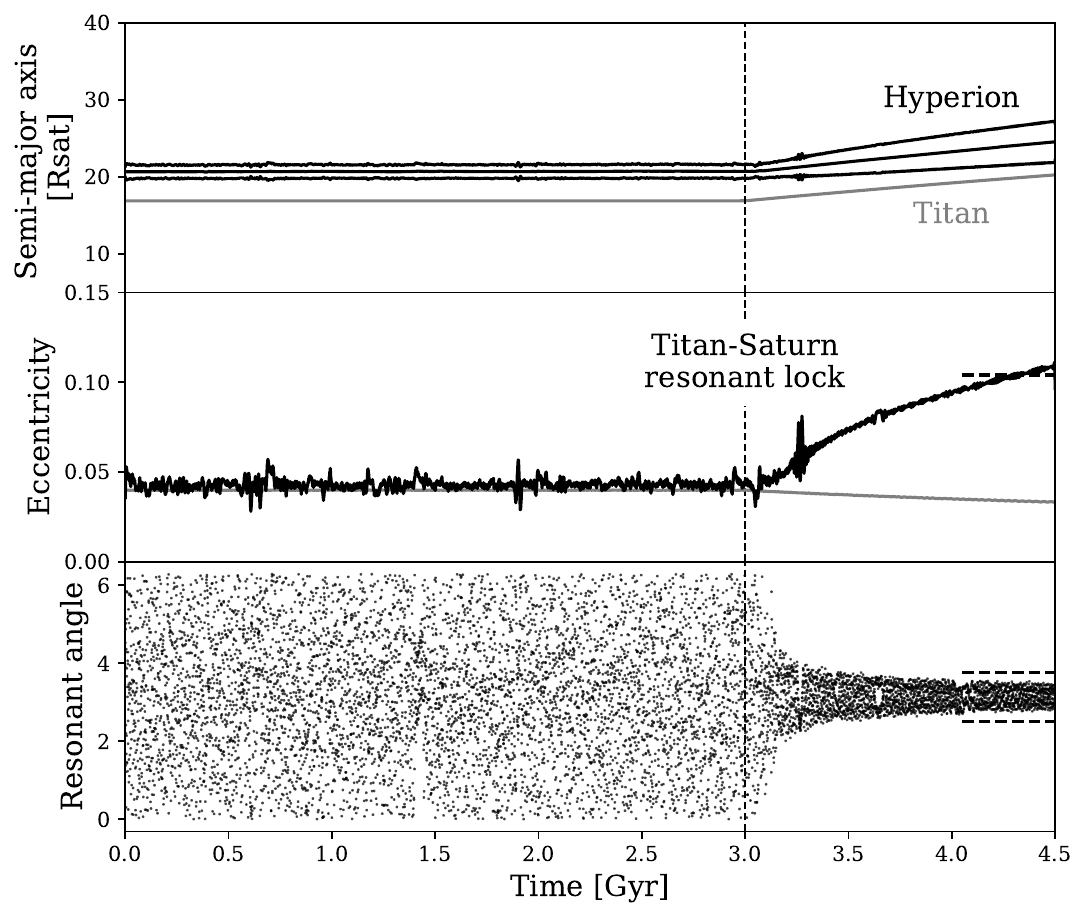}
    \caption{Capture of Hyperion into 4:3 resonance with Titan in the migration-by-resonant locking model. In both panels we have assumed $Q_\text{H}=40$, on the left $t_\text{lock}=\qty{1}{Gyr}$ and on the right $t_\text{lock}=\qty{3}{Gyr}$. The resonant angle plotted in the bottom panel is $\phi_2=4\lambda_\text{H}-3\lambda_\text{Ti}-\varpi_\text{H}$. Dashed lines in the bottom two rows show the measured values of Hyperion's (forced) eccentricity and libration angle range.}
    \label{fig:hyp_capture_lock}
\end{figure}

The simulations consistently capture Hyperion into the observed 4:3 resonant configuration with $e_H\approx 0.1$ if $Q_\text{H}\approx 40$. The final eccentricity does not depend strongly on $t_\text{lock}$.
Figure \ref{fig:hyp_capture_lock} shows two of these integrations where we have used $Q_\text{H}=40$ and $t_\text{lock}=\qty{1}{Gyr}$ (left panel) and $t_\text{lock}=\qty{3}{Gyr}$ (right panel). In both cases, Hyperion successfully captures into the 4:3 mean motion resonance with Titan after Titan begins migrating outward. Initially, Hyperion's eccentricity is suppressed to Titan's eccentricity, to which it is secularly coupled, by the dissipation resulting from chaotic tumbling. Once Hyperion reaches a sufficient semi-major axis, the resonant excitation from Titan becomes stronger than the tidal damping and Hyperion becomes more eccentric. By the end of the simulation, Hyperion has reached an eccentricity close its present value of $0.1$. The amplitude of libration of the resonant angle is significant, even if somewhat smaller than what is observed.

Critically, this model of Hyperion's capture into resonance is not compatible with long-range Titan migration if damping in Hyperion is ignored. In such an undamped scenario, to prevent Hyperion from having too large of an eccentricity, Titan must only migrate $4\%$ in semi-major axis (i.e.~$t_\text{lock} \approx \qty{4.0}{Gyr}$) after the 4:3 capture. If, however, Titan migrated more than $11\%$ from its initial location (i.e.~$t_\text{lock}\lesssim\qty{3.2}{Gyr}$), Titan and Hyperion would have started wide of, and then encountered, the 3:2 resonance. The encounter is adiabatic \citep{Batygin2015} and capture into the 3:2 is almost guaranteed, unless the eccentricity of Hyperion is very large.\footnote{\cite{Colombo1974} argue, using a backwards integration, that Hyperion would have avoided capture into the 2:1 and 3:2 resonances, but their reasoning is flawed. Because resonant encounters in the backwards integration are divergent, permanent} capture into resonance is impossible regardless of migration speed \citep{Henrard1982}. Once in the wrong resonance, Hyperion will grow in eccentricity and eventually be ejected, never entering the 4:3 resonance. Thus, at face value, any model in which Titan and Hyperion cross first-order resonances adiabatically requires that Hyperion and Titan must have started interior to the 3:2 resonance. However, incorporating tidal dissipation in Hyperion removes the fine-tuning restriction that the resonant lock-driven migration of Titan can only have begun recently.

\subsubsection{Slower Titan Migration}
\label{sec:slow}
Now, we consider the measurement of Jacobson (2022), who find $Q_\mSaturn(n_\text{Ti})=1224\pm 119$, or $\tau_{a,\text{Ti}}\approx -\qty{100}{Gyr}$. Before proceeding, we remark that in the context of this measurement, it is not obvious what the source of dissipation with Saturn would be. \cite{Cuk2023} argue that $Q_\mSaturn \approx 1200$ is the frequency-independent dissipation within Saturn and that the \cite{Jacobson2022} measurement would imply that Titan is experiencing equilibrium tides outside a resonant lock. However, the migration of Tethys implies $Q_\mSaturn \approx 7000$ \citep{Lainey2020} and it is not clear how one moon could experience tidal dissipation in Saturn weaker than equilibrium. In the absence of a clear guide, we take $Q_\mSaturn=1200$ and assume $Q_\mSaturn$ is constant over time and forcing frequency. Integrating Eq.~\ref{eq:tit_tid} from $t=0$ to $t=\qty{4.5}{Gyr}$, we find that with these assumptions, Titan's orbit has expanded by $6.1\%$ over the age of the Solar System, so avoiding capture of Hyperion into the 3:2 resonance is not a concern.

\begin{figure}
    \centering
    \includegraphics[width=0.47\textwidth]{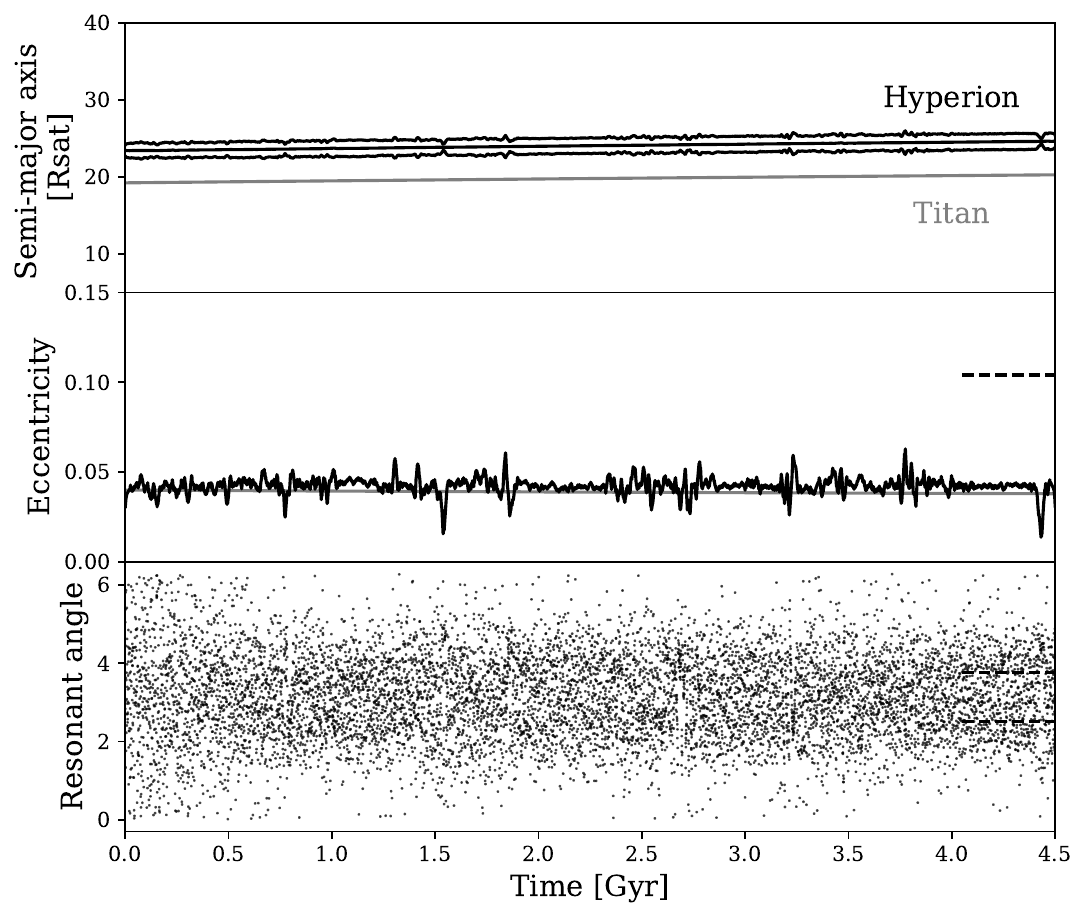}
    \includegraphics[width=0.47\textwidth]{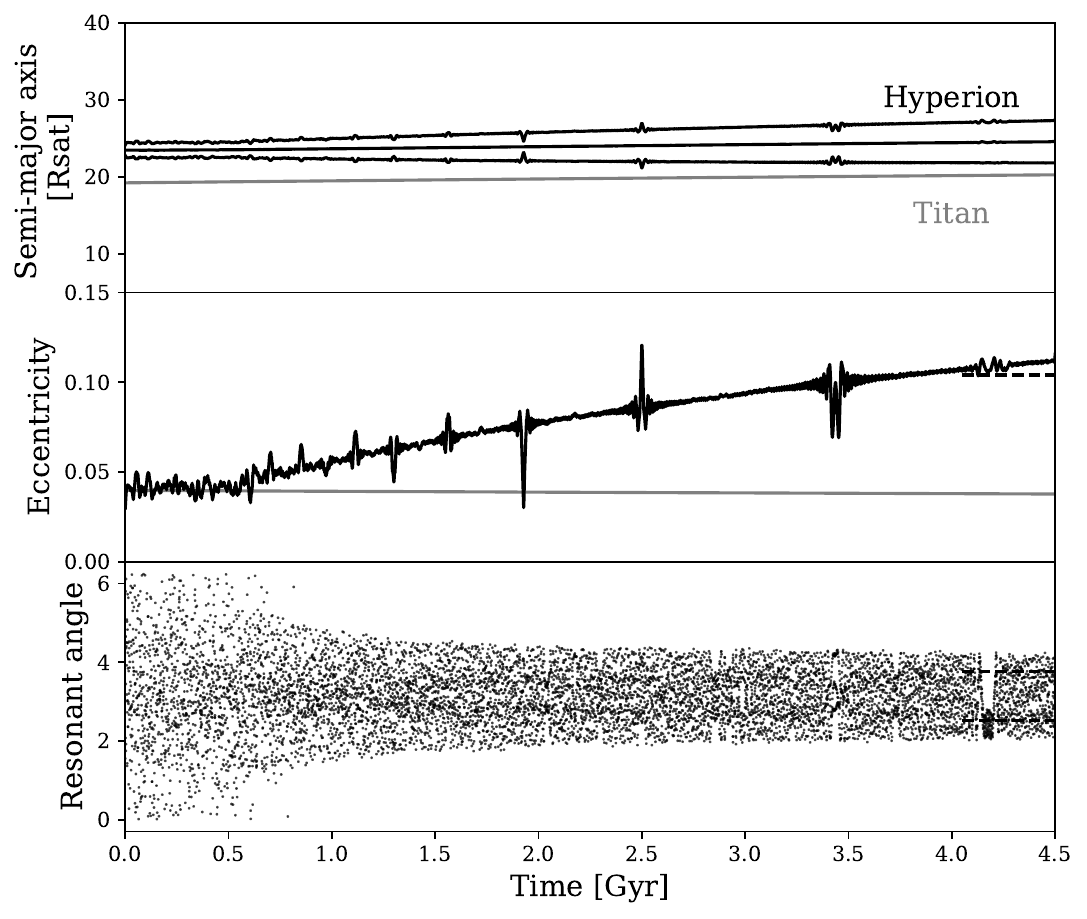}
    \caption{Capture of Hyperion into 4:3 resonance with Titan in the equilibrium tides model where $Q_\mSaturn = 1200$. On the left panel we have assumed $Q_\text{H}=100$, and on the right $Q_\text{H}=\infty$, corresponding to no damping in Hyperion. Dashed lines in the bottom two rows show the measured values of Hyperion's (forced) eccentricity and libration angle range.}
    \label{fig:hyp_capture_Q}
\end{figure}

We ran another suite of simulations with this model of slower Titan migration. Titan was initialized with an initial semi-major axis of $19.2 R_\mSaturn$ and eccentricity of 0.04. Migration of Titan was computed with Eq.~\ref{eq:tit_tid} and eccentricity damping was assumed, as in Section~\ref{sec:fast}, to be $\tau_{e,\text{Ti}}=\tau_{a,\text{Ti}}$. The strength of dissipation in Saturn was set to $Q_\mSaturn/k_{2,\mSaturn}=3000$, so that after $\qty{4.5}{Gyr}$ Titan would reach its current semi-major axis. Hyperion was initialized with several period ratios between the 4:3 and 3:2 resonance with Titan, and several eccentricities from 0 to 0.05.

Figure \ref{fig:hyp_capture_Q} shows two simulations of capture in this model of Titan migration, with $Q_\text{H}=100$ (left panel) and $Q_\text{H}=\infty$ (right panel). 
The first value is generally expected for rocky bodies \citep{Goldreich2009,Brasser2020} and puts Hyperion in the regime where tidal dissipation in Hyperion dominates over resonant excitation of eccentricity from Titan's slower migration. Hyperion remains secularly coupled to Titan and the resonant angle circulates. Conversely, $Q_\text{H}=\infty$ corresponds to no dissipation in Hyperion and is equivalent to the tidal capture hypothesis of \cite{Colombo1974} and \cite{Cuk2013}. In this case, Hyperion reaches its present eccentricity and libration angle amplitude at the end of the simulation. Using the range of parameters in our simulation suite, we find that Hyperion only reaches its current eccentricity of $0.1$ if $Q_\text{H}\gtrsim 1000$. 

Figures \ref{fig:hyp_capture_lock} and \ref{fig:hyp_capture_Q} demonstrate that slower Titan migration demands much weaker tidal dissipation in Hyperion. In the rapid migration scenario (Figure \ref{fig:hyp_capture_lock}), eccentricity pumping from resonant excitation is roughly comparable to damping from tidal dissipation for $Q_\text{H}\sim 100$, allowing the eccentricity to gradually grow to its present value. In contrast, resonant excitation from a slowly migrating Titan (Figure \ref{fig:hyp_capture_Q}) is dwarfed by tidal dissipation unless $Q_\text{H}$ is quite large.

This sort of weak dissipation in Hyperion is not physically implausible. \cite{Nimmo2019} argue that in rocky rubble pile asteroids, energy losses occur only in a thin surface layer of regolith rather than the entire body. In the context of that model, they find that $Q$ scales as $R^2$ and $Q/k_2$ as $R$. Hyperion is icy and much larger than the typical rubble pile asteroids they investigate, but extrapolating their model would predict that Hyperion has a very large $Q/k_2$.

\section{Discussion and Conclusions}
\label{sec:discussion}

Despite its apparent simplicity, the outer saturnian system has confounded understanding for decades and arguably has become more challenging with new discoveries. At face value, the orbit of Hyperion implies that Titan migrated slowly, while the survival of Iapetus requires faster migration. Both cases rely on tidal dissipation in Saturn much stronger than conventionally expected. Taken together, these properties are a challenge to reconcile, especially while remaining consistent with constraints provided by the inner system.

We have shown that Hyperion's rotation is the key to resolving this discrepancy. Hyperion's spin rate depends steeply on its orbital eccentricity, a consequence of a chaotic sea generated from the overlap of spin-orbit resonances that grows with eccentricity. Tidal dissipation is much stronger at higher spins and is thus a steep function of $e_\text{H}$ as well. Although the mean-motion resonant interaction between Titan and Hyperion grows the latter's orbital eccentricity as Titan migrates outward, the enhanced tidal dissipation resulting from rapid tumbling  damps the eccentricity. The degree of dissipation within Hyperion can be probed by comparing the relative strengths of these two effects to match the observed $e_\text{H}=0.1$. The precise rate of Titan's migration is disputed in recent works. If it is rapid, Hyperion must have a tidal quality factor of $Q_\text{H}\approx 40$, similar to what is typically expected for rocky bodies. Alternatively, if migration is slow, Hyperion must be only weakly dissipative ($Q_\text{H}\gtrsim 1000$), which is reasonable if dissipation occurs solely in a surface layer, as suggested by \cite{Nimmo2019}.

The dissipation itself could in principle be detected directly through an excess thermal signature. Energy provided by the orbit acts to heat up Hyperion and sublimate its water ice. If Titan is migrating rapidly and $Q_\text{H}=40$ (e.g. right panel of Figure \ref{fig:hyp_capture_lock}), the current energy dissipation rate is $dE/dt\sim \qty{3}{MW}$. As a crude approximation, if this dissipation were constant over the lifetime of the Solar System, and assuming Hyperion is made entirely of water ice, approximately 3\% of the mass of Hyperion would have sublimated due to the tidal dissipation. While non-negligible, this amount is insufficient to explain Hyperion's high ($>40\%$) internal porosity \citep{Thomas2007}. However, it is possible that Hyperion experienced a transient phase of high eccentricity, perhaps due to a scattering event before the resonant capture with Titan. An excitation to $e_\text{H}\approx 0.3$ followed by damping to a circular orbit could sublimate $\approx 40\%$ of Hyperion's mass in $<\qty{1}{Gyr}$ and account for its current porosity.

Significant sublimation could have other impacts. \citet{Seligman2020} demonstrate that uniform sublimation across the surface of an ellipsoid acts to elongate it; they use this effect to explain the extreme body axis ratio of `Oumuamua. If Hyperion underwent a similar process, its current shape could therefore be the consequence of a small irregularity in the shape of the primordial Hyperion that grew as material preferentially sublimated from certain regions.

These two complications highlight an important assumption we have made throughout this work. We have taken Hyperion's shape and material properties to be constant over the lifetime of the Solar System. However, sublimation, gravitational settling from tumbling, and impacts can vary the mass, composition, porosity, strength, and shape of Hyperion, all of which would affect the tidal dissipation rate. Incorporating all of these effects, while challenging, would provide a more complete picture of Hyperion's evolution.

Our detailed investigations of Hyperion's rotation showed that it is more complex than the original predictions derived from a one-dimensional model. Even so, generation of chaos via overlap of non-linear spin-orbit resonances remains qualitatively useful. Our results suggest that Hyperion has had a rich rotational history, alternating between tumbling and quasi-regular states that depend on its instantaneous eccentricity. The orbit and rotation of Hyperion are inextricably coupled: orbital eccentricity sets the typical spin rate, and in turn, the nonsynchronous spin damps the orbital eccentricity. Similar spin-to-orbit and orbit-to-spin coupling has been suggested for asteroid binaries \citep{Efroimsky2015,Nimmo2019,Quillen2022} and close-in satellites \citep{Dobrovolskis1997,Quillen2017,Quillen2020}. Hyperion, despite its small size and great distance from Saturn, is subject to the same complex feedback between spin and orbital evolution.

Although we have identified a compelling \textit{process} to explain the current state of Hyperion, we have not determined the exact \textit{scenario} that transpired. Of particular interest is Hyperion's initial orbit. If its original period ratio with Titan exceeded 1.5, capture into the 3:2 resonance is highly likely, unless another mechanism can break the resonance or avoid capture entirely. On the other hand, formation of Hyperion in such close proximity to the very massive Titan seems \textit{a priori} unlikely. We encourage further work on this topic. Finally, our rotation model of Hyperion does not include the effects of tidal despinning. While it is clear that the current rotation state is not in the 1:1 or 3:2 island, it remains possible that it lies on a strange attractor with perpetual chaotic but quasi-regular motion \citep{Melnikov2014}. A self-consistent rotation model incorporating wobble damping and tidal dissipation that is coupled to the orbital evolution would thus be necessary to further constrain the history of Hyperion and more precisely measure its tidal parameters.

\section*{Acknowledgements}
We thank the two referees for thorough reading of the manuscript and helpful feedback. We also thank David Nesvorn\'{y}, Rogerio Deienno, Bill Bottke, and Jim Fuller for insightful discussions. K.B. is grateful to Caltech, the David and Lucile Packard Foundation, and National Science Foundation (grant number: AST 2109276) for their generous support.

\bibliographystyle{cas-model2-names}

\bibliography{main}

\begin{thebibliography}{52}
\expandafter\ifx\csname natexlab\endcsname\relax\def\natexlab#1{#1}\fi
\providecommand{\url}[1]{\texttt{#1}}
\providecommand{\href}[2]{#2}
\providecommand{\path}[1]{#1}
\providecommand{\DOIprefix}{doi:}
\providecommand{\ArXivprefix}{arXiv:}
\providecommand{\URLprefix}{URL: }
\providecommand{\Pubmedprefix}{pmid:}
\providecommand{\doi}[1]{\href{http://dx.doi.org/#1}{\path{#1}}}
\providecommand{\Pubmed}[1]{\href{pmid:#1}{\path{#1}}}
\providecommand{\bibinfo}[2]{#2}
\ifx\xfnm\relax \def\xfnm[#1]{\unskip,\space#1}\fi
\bibitem[{Agrusa et~al.(2021)Agrusa, Gkolias, Tsiganis, Richardson, Meyer,
  Scheeres, {\'C}uk, Jacobson, Michel, Karatekin, Cheng, Hirabayashi, Zhang,
  Fahnestock and Davis}]{Agrusa2021}
\bibinfo{author}{Agrusa, H.F.}, \bibinfo{author}{Gkolias, I.},
  \bibinfo{author}{Tsiganis, K.}, \bibinfo{author}{Richardson, D.C.},
  \bibinfo{author}{Meyer, A.J.}, \bibinfo{author}{Scheeres, D.J.},
  \bibinfo{author}{{\'C}uk, M.}, \bibinfo{author}{Jacobson, S.A.},
  \bibinfo{author}{Michel, P.}, \bibinfo{author}{Karatekin, {\"O}.},
  \bibinfo{author}{Cheng, A.F.}, \bibinfo{author}{Hirabayashi, M.},
  \bibinfo{author}{Zhang, Y.}, \bibinfo{author}{Fahnestock, E.G.},
  \bibinfo{author}{Davis, A.B.}, \bibinfo{year}{2021}.
\newblock \bibinfo{title}{The excited spin state of {{Dimorphos}} resulting
  from the {{DART}} impact}.
\newblock \bibinfo{journal}{Icarus} \bibinfo{volume}{370},
  \bibinfo{pages}{114624}.
\newblock \DOIprefix\doi{10.1016/j.icarus.2021.114624}.
\bibitem[{Arribas et~al.(2006)Arribas, Elipe and Palacios}]{Arribas2006}
\bibinfo{author}{Arribas, M.}, \bibinfo{author}{Elipe, A.},
  \bibinfo{author}{Palacios, M.}, \bibinfo{year}{2006}.
\newblock \bibinfo{title}{Quaternions and the rotation of a rigid body}.
\newblock \bibinfo{journal}{Celestial Mech Dyn Astr} \bibinfo{volume}{96},
  \bibinfo{pages}{239--251}.
\newblock \DOIprefix\doi{10.1007/s10569-006-9037-6}.
\bibitem[{Batygin(2015)}]{Batygin2015}
\bibinfo{author}{Batygin, K.}, \bibinfo{year}{2015}.
\newblock \bibinfo{title}{Capture of planets into mean-motion resonances and
  the origins of extrasolar orbital architectures}.
\newblock \bibinfo{journal}{Monthly Notices of the Royal Astronomical Society}
  \bibinfo{volume}{451}, \bibinfo{pages}{2589--2609}.
\newblock \DOIprefix\doi{10.1093/mnras/stv1063}.
\bibitem[{Batygin et~al.(2023)Batygin, Adams and Becker}]{Batygin2023b}
\bibinfo{author}{Batygin, K.}, \bibinfo{author}{Adams, F.C.},
  \bibinfo{author}{Becker, J.}, \bibinfo{year}{2023}.
\newblock \bibinfo{title}{The {{Origin}} of {{Universality}} in the {{Inner
  Edges}} of {{Planetary Systems}}}.
\newblock \bibinfo{journal}{The Astrophysical Journal} \bibinfo{volume}{951},
  \bibinfo{pages}{L19}.
\newblock \DOIprefix\doi{10.3847/2041-8213/acdb5d}.
\bibitem[{Black et~al.(1995)Black, Nicholson and Thomas}]{Black1995}
\bibinfo{author}{Black, G.J.}, \bibinfo{author}{Nicholson, P.D.},
  \bibinfo{author}{Thomas, P.C.}, \bibinfo{year}{1995}.
\newblock \bibinfo{title}{Hyperion: {{Rotational}} dynamics.}
\newblock \bibinfo{journal}{Icarus} \bibinfo{volume}{117},
  \bibinfo{pages}{149--161}.
\newblock \DOIprefix\doi{10.1006/icar.1995.1148}.
\bibitem[{Bottke et~al.(2023)Bottke, Vokrouhlick{\'y}, Marshall, Nesvorn{\'y},
  Morbidelli, Deienno, Marchi, Dones and Levison}]{Bottke2023}
\bibinfo{author}{Bottke, W.F.}, \bibinfo{author}{Vokrouhlick{\'y}, D.},
  \bibinfo{author}{Marshall, R.}, \bibinfo{author}{Nesvorn{\'y}, D.},
  \bibinfo{author}{Morbidelli, A.}, \bibinfo{author}{Deienno, R.},
  \bibinfo{author}{Marchi, S.}, \bibinfo{author}{Dones, L.},
  \bibinfo{author}{Levison, H.F.}, \bibinfo{year}{2023}.
\newblock \bibinfo{title}{The {{Collisional Evolution}} of the {{Primordial
  Kuiper Belt}}, {{Its Destabilized Population}}, and the {{Trojan
  Asteroids}}}.
\newblock \bibinfo{journal}{The Planetary Science Journal} \bibinfo{volume}{4},
  \bibinfo{pages}{168}.
\newblock \DOIprefix\doi{10.3847/PSJ/ace7cd}.
\bibitem[{Brasser(2020)}]{Brasser2020}
\bibinfo{author}{Brasser, R.}, \bibinfo{year}{2020}.
\newblock \bibinfo{title}{Efficient tidal dissipation in {{Deimos}}}.
\newblock \bibinfo{journal}{Icarus} \bibinfo{volume}{347},
  \bibinfo{pages}{113791}.
\newblock \DOIprefix\doi{10.1016/j.icarus.2020.113791}.
\bibitem[{Burns and Safronov(1973)}]{Burns1973}
\bibinfo{author}{Burns, J.A.}, \bibinfo{author}{Safronov, V.S.},
  \bibinfo{year}{1973}.
\newblock \bibinfo{title}{Asteroid nutation angles}.
\newblock \bibinfo{journal}{Monthly Notices of the Royal Astronomical Society}
  \bibinfo{volume}{165}, \bibinfo{pages}{403}.
\newblock \DOIprefix\doi{10.1093/mnras/165.4.403}.
\bibitem[{Chirikov(1979)}]{Chirikov1979}
\bibinfo{author}{Chirikov, B.V.}, \bibinfo{year}{1979}.
\newblock \bibinfo{title}{A universal instability of many-dimensional
  oscillator systems}.
\newblock \bibinfo{journal}{Physics Reports} \bibinfo{volume}{52},
  \bibinfo{pages}{263--379}.
\newblock \DOIprefix\doi{10.1016/0370-1573(79)90023-1}.
\bibitem[{Colombo et~al.(1974)Colombo, Franklin and Shapiro}]{Colombo1974}
\bibinfo{author}{Colombo, G.}, \bibinfo{author}{Franklin, F.A.},
  \bibinfo{author}{Shapiro, I.I.}, \bibinfo{year}{1974}.
\newblock \bibinfo{title}{On the formation of the orbit-orbit resonance of
  {{Titan}} and {{Hyperion}}}.
\newblock \bibinfo{journal}{The Astronomical Journal} \bibinfo{volume}{79},
  \bibinfo{pages}{61}.
\newblock \DOIprefix\doi{10.1086/111533}.
\bibitem[{{\'C}uk et~al.(2013){\'C}uk, Dones and Nesvorn{\'y}}]{Cuk2013}
\bibinfo{author}{{\'C}uk, M.}, \bibinfo{author}{Dones, L.},
  \bibinfo{author}{Nesvorn{\'y}, D.}, \bibinfo{year}{2013}.
\newblock \bibinfo{title}{Titan-{{Hyperion Resonance}} and the {{Tidal Q}} of
  {{Saturn}}}.
\bibitem[{{\'C}uk et~al.(2016a){\'C}uk, Dones and Nesvorn{\'y}}]{Cuk2016}
\bibinfo{author}{{\'C}uk, M.}, \bibinfo{author}{Dones, L.},
  \bibinfo{author}{Nesvorn{\'y}, D.}, \bibinfo{year}{2016}a.
\newblock \bibinfo{title}{Dynamical {{Evidence}} for a {{Late Formation}} of
  {{Saturn}}'s {{Moons}}}.
\newblock \bibinfo{journal}{The Astrophysical Journal} \bibinfo{volume}{820},
  \bibinfo{pages}{97}.
\newblock \DOIprefix\doi{10.3847/0004-637X/820/2/97}.
\bibitem[{{\'C}uk and El~Moutamid(2023)}]{Cuk2023}
\bibinfo{author}{{\'C}uk, M.}, \bibinfo{author}{El~Moutamid, M.},
  \bibinfo{year}{2023}.
\newblock \bibinfo{title}{A {{Past Episode}} of {{Rapid Tidal Evolution}} of
  {{Enceladus}}?}
\newblock \bibinfo{journal}{The Planetary Science Journal} \bibinfo{volume}{4},
  \bibinfo{pages}{119}.
\newblock \DOIprefix\doi{10.3847/PSJ/acde80}.
\bibitem[{{\'C}uk et~al.(2016b){\'C}uk, Hamilton, Lock and Stewart}]{Cuk2016a}
\bibinfo{author}{{\'C}uk, M.}, \bibinfo{author}{Hamilton, D.P.},
  \bibinfo{author}{Lock, S.J.}, \bibinfo{author}{Stewart, S.T.},
  \bibinfo{year}{2016}b.
\newblock \bibinfo{title}{Tidal evolution of the {{Moon}} from a
  high-obliquity, high-angular-momentum {{Earth}}}.
\newblock \bibinfo{journal}{Nature} \bibinfo{volume}{539},
  \bibinfo{pages}{402--406}.
\newblock \DOIprefix\doi{10.1038/nature19846}.
\bibitem[{Dobrovolskis(1995)}]{Dobrovolskis1995}
\bibinfo{author}{Dobrovolskis, A.R.}, \bibinfo{year}{1995}.
\newblock \bibinfo{title}{Chaotic {{Rotation}} of {{Nereid}}?}
\newblock \bibinfo{journal}{Icarus} \bibinfo{volume}{118},
  \bibinfo{pages}{181--198}.
\newblock \DOIprefix\doi{10.1006/icar.1995.1184}.
\bibitem[{Dobrovolskis et~al.(1997)Dobrovolskis, Peale and
  Harris}]{Dobrovolskis1997}
\bibinfo{author}{Dobrovolskis, A.R.}, \bibinfo{author}{Peale, S.J.},
  \bibinfo{author}{Harris, A.W.}, \bibinfo{year}{1997}.
\newblock \bibinfo{title}{Dynamics of the {{Pluto-Charon Binary}}}.
\bibitem[{Efroimsky(2015)}]{Efroimsky2015}
\bibinfo{author}{Efroimsky, M.}, \bibinfo{year}{2015}.
\newblock \bibinfo{title}{Tidal {{Evolution}} of {{Asteroidal Binaries}}.
  {{Ruled}} by {{Viscosity}}. {{Ignorant}} of {{Rigidity}}.}
\newblock \bibinfo{journal}{The Astronomical Journal} \bibinfo{volume}{150},
  \bibinfo{pages}{98}.
\newblock \DOIprefix\doi{10.1088/0004-6256/150/4/98}.
\bibitem[{Fuller et~al.(2016)Fuller, Luan and Quataert}]{Fuller2016}
\bibinfo{author}{Fuller, J.}, \bibinfo{author}{Luan, J.},
  \bibinfo{author}{Quataert, E.}, \bibinfo{year}{2016}.
\newblock \bibinfo{title}{Resonance locking as the source of rapid tidal
  migration in the {{Jupiter}} and {{Saturn}} moon systems}.
\newblock \bibinfo{journal}{Monthly Notices of the Royal Astronomical Society}
  \bibinfo{volume}{458}, \bibinfo{pages}{3867--3879}.
\newblock \DOIprefix\doi{10.1093/mnras/stw609}.
\bibitem[{Goldreich and Peale(1966)}]{Goldreich1966}
\bibinfo{author}{Goldreich, P.}, \bibinfo{author}{Peale, S.},
  \bibinfo{year}{1966}.
\newblock \bibinfo{title}{Spin-orbit coupling in the solar system}.
\newblock \bibinfo{journal}{The Astronomical Journal} \bibinfo{volume}{71},
  \bibinfo{pages}{425}.
\newblock \DOIprefix\doi{10.1086/109947}.
\bibitem[{Goldreich and Sari(2009)}]{Goldreich2009}
\bibinfo{author}{Goldreich, P.}, \bibinfo{author}{Sari, R.},
  \bibinfo{year}{2009}.
\newblock \bibinfo{title}{Tidal {{Evolution}} of {{Rubble Piles}}}.
\newblock \bibinfo{journal}{The Astrophysical Journal} \bibinfo{volume}{691},
  \bibinfo{pages}{54--60}.
\newblock \DOIprefix\doi{10.1088/0004-637X/691/1/54}.
\bibitem[{Goldreich and Soter(1966)}]{Goldreich1966a}
\bibinfo{author}{Goldreich, P.}, \bibinfo{author}{Soter, S.},
  \bibinfo{year}{1966}.
\newblock \bibinfo{title}{Q in the {{Solar System}}}.
\newblock \bibinfo{journal}{Icarus} \bibinfo{volume}{5},
  \bibinfo{pages}{375--389}.
\newblock \DOIprefix\doi{10.1016/0019-1035(66)90051-0}.
\bibitem[{Harbison et~al.(2011)Harbison, Thomas and Nicholson}]{Harbison2011}
\bibinfo{author}{Harbison, R.A.}, \bibinfo{author}{Thomas, P.C.},
  \bibinfo{author}{Nicholson, P.C.}, \bibinfo{year}{2011}.
\newblock \bibinfo{title}{Rotational modeling of {{Hyperion}}}.
\newblock \bibinfo{journal}{Celestial Mechanics and Dynamical Astronomy}
  \bibinfo{volume}{110}, \bibinfo{pages}{1--16}.
\newblock \DOIprefix\doi{10.1007/s10569-011-9337-3}.
\bibitem[{Henrard(1982)}]{Henrard1982}
\bibinfo{author}{Henrard, J.}, \bibinfo{year}{1982}.
\newblock \bibinfo{title}{Capture into resonance - an extension of the use of
  adiabatic invariants}.
\newblock \bibinfo{journal}{Celestial Mechanics} \bibinfo{volume}{27},
  \bibinfo{pages}{3--22}.
\newblock \DOIprefix\doi{10.1007/BF01228946}.
\bibitem[{Jacobson(2022)}]{Jacobson2022}
\bibinfo{author}{Jacobson, {\relax Robert}.A.}, \bibinfo{year}{2022}.
\newblock \bibinfo{title}{The {{Orbits}} of the {{Main Saturnian Satellites}},
  the {{Saturnian System Gravity Field}}, and the {{Orientation}} of
  {{Saturn}}'s {{Pole}}}.
\newblock \bibinfo{journal}{The Astronomical Journal} \bibinfo{volume}{164},
  \bibinfo{pages}{199}.
\newblock \DOIprefix\doi{10.3847/1538-3881/ac90c9}.
\bibitem[{Karney(1983)}]{Karney1983}
\bibinfo{author}{Karney, C.F.F.}, \bibinfo{year}{1983}.
\newblock \bibinfo{title}{Long-time correlations in the stochastic regime}.
\newblock \bibinfo{journal}{Physica D: Nonlinear Phenomena}
  \bibinfo{volume}{8}, \bibinfo{pages}{360--380}.
\newblock \DOIprefix\doi{10.1016/0167-2789(83)90232-4}.
\bibitem[{Klavetter(1989)}]{Klavetter1989}
\bibinfo{author}{Klavetter, J.J.}, \bibinfo{year}{1989}.
\newblock \bibinfo{title}{Rotation of {{Hyperion}}. {{II}}. {{Dynamics}}}.
\newblock \bibinfo{journal}{The Astronomical Journal} \bibinfo{volume}{98},
  \bibinfo{pages}{1855}.
\newblock \DOIprefix\doi{10.1086/115264}.
\bibitem[{Lainey et~al.(2020)Lainey, Casajus, Fuller, Zannoni, Tortora, Cooper,
  Murray, Modenini, Park, Robert and Zhang}]{Lainey2020}
\bibinfo{author}{Lainey, V.}, \bibinfo{author}{Casajus, L.G.},
  \bibinfo{author}{Fuller, J.}, \bibinfo{author}{Zannoni, M.},
  \bibinfo{author}{Tortora, P.}, \bibinfo{author}{Cooper, N.},
  \bibinfo{author}{Murray, C.}, \bibinfo{author}{Modenini, D.},
  \bibinfo{author}{Park, R.S.}, \bibinfo{author}{Robert, V.},
  \bibinfo{author}{Zhang, Q.}, \bibinfo{year}{2020}.
\newblock \bibinfo{title}{Resonance locking in giant planets indicated by the
  rapid orbital expansion of {{Titan}}}.
\newblock \bibinfo{journal}{Nature Astronomy} \bibinfo{volume}{4},
  \bibinfo{pages}{1053--1058}.
\newblock \DOIprefix\doi{10.1038/s41550-020-1120-5}.
\bibitem[{Lainey et~al.(2017)Lainey, Jacobson, Tajeddine, Cooper, Murray,
  Robert, Tobie, Guillot, Mathis, Remus, Desmars, Arlot, De~Cuyper, Dehant,
  Pascu, Thuillot, {Le Poncin-Lafitte} and Zahn}]{Lainey2017}
\bibinfo{author}{Lainey, V.}, \bibinfo{author}{Jacobson, R.A.},
  \bibinfo{author}{Tajeddine, R.}, \bibinfo{author}{Cooper, N.J.},
  \bibinfo{author}{Murray, C.}, \bibinfo{author}{Robert, V.},
  \bibinfo{author}{Tobie, G.}, \bibinfo{author}{Guillot, T.},
  \bibinfo{author}{Mathis, S.}, \bibinfo{author}{Remus, F.},
  \bibinfo{author}{Desmars, J.}, \bibinfo{author}{Arlot, J.E.},
  \bibinfo{author}{De~Cuyper, J.P.}, \bibinfo{author}{Dehant, V.},
  \bibinfo{author}{Pascu, D.}, \bibinfo{author}{Thuillot, W.},
  \bibinfo{author}{{Le Poncin-Lafitte}, C.}, \bibinfo{author}{Zahn, J.P.},
  \bibinfo{year}{2017}.
\newblock \bibinfo{title}{New constraints on {{Saturn}}'s interior from
  {{Cassini}} astrometric data}.
\newblock \bibinfo{journal}{Icarus} \bibinfo{volume}{281},
  \bibinfo{pages}{286--296}.
\newblock \DOIprefix\doi{10.1016/j.icarus.2016.07.014}.
\bibitem[{Lainey et~al.(2012)Lainey, Karatekin, Desmars, Charnoz, Arlot,
  Emelyanov, {Le Poncin-Lafitte}, Mathis, Remus, Tobie and Zahn}]{Lainey2012}
\bibinfo{author}{Lainey, V.}, \bibinfo{author}{Karatekin, {\"O}.},
  \bibinfo{author}{Desmars, J.}, \bibinfo{author}{Charnoz, S.},
  \bibinfo{author}{Arlot, J.E.}, \bibinfo{author}{Emelyanov, N.},
  \bibinfo{author}{{Le Poncin-Lafitte}, C.}, \bibinfo{author}{Mathis, S.},
  \bibinfo{author}{Remus, F.}, \bibinfo{author}{Tobie, G.},
  \bibinfo{author}{Zahn, J.P.}, \bibinfo{year}{2012}.
\newblock \bibinfo{title}{Strong {{Tidal Dissipation}} in {{Saturn}} and
  {{Constraints}} on {{Enceladus}}' {{Thermal State}} from {{Astrometry}}}.
\newblock \bibinfo{journal}{The Astrophysical Journal} \bibinfo{volume}{752},
  \bibinfo{pages}{14}.
\newblock \DOIprefix\doi{10.1088/0004-637X/752/1/14}.
\bibitem[{Meiss(1992)}]{Meiss1992}
\bibinfo{author}{Meiss, J.D.}, \bibinfo{year}{1992}.
\newblock \bibinfo{title}{Symplectic maps, variational principles, and
  transport}.
\newblock \bibinfo{journal}{Reviews of Modern Physics} \bibinfo{volume}{64},
  \bibinfo{pages}{795--848}.
\newblock \DOIprefix\doi{10.1103/RevModPhys.64.795}.
\bibitem[{Melnikov(2014)}]{Melnikov2014}
\bibinfo{author}{Melnikov, A.V.}, \bibinfo{year}{2014}.
\newblock \bibinfo{title}{Conditions for appearance of strange attractors in
  rotational dynamics of small planetary satellites}.
\newblock \bibinfo{journal}{Cosmic Research} \bibinfo{volume}{52},
  \bibinfo{pages}{461--471}.
\newblock \DOIprefix\doi{10.1134/S0010952514060045}.
\bibitem[{Mel'nikov(2020)}]{Melnikov2020}
\bibinfo{author}{Mel'nikov, A.V.}, \bibinfo{year}{2020}.
\newblock \bibinfo{title}{Orientation of {{Figures}} of {{Small Planetary
  Satellites During Chaotic Rotation}}}.
\newblock \bibinfo{journal}{Solar System Research} \bibinfo{volume}{54},
  \bibinfo{pages}{432--441}.
\newblock \DOIprefix\doi{10.1134/S0038094620050068}.
\bibitem[{Morbidelli(2002)}]{Morbidelli2002}
\bibinfo{author}{Morbidelli, A.}, \bibinfo{year}{2002}.
\newblock \bibinfo{title}{Modern Celestial Mechanics : Aspects of Solar System
  Dynamics}.
\newblock \bibinfo{publisher}{{Taylor \& Francis}}.
\bibitem[{Murray and Dermott(1999)}]{Murray1999}
\bibinfo{author}{Murray, C.D.}, \bibinfo{author}{Dermott, S.F.},
  \bibinfo{year}{1999}.
\newblock \bibinfo{title}{Solar system dynamics}.
\newblock \bibinfo{journal}{Solar system dynamics by C.D. Murray and S.F.
  McDermott. (Cambridge, UK: Cambridge University Press), ISBN 0-521-57295-9
  (hc.), ISBN 0-521-57297-4 (pbk.).} .
\bibitem[{Nimmo and Matsuyama(2019)}]{Nimmo2019}
\bibinfo{author}{Nimmo, F.}, \bibinfo{author}{Matsuyama, I.},
  \bibinfo{year}{2019}.
\newblock \bibinfo{title}{Tidal dissipation in rubble-pile asteroids}.
\newblock \bibinfo{journal}{Icarus} \bibinfo{volume}{321},
  \bibinfo{pages}{715--721}.
\newblock \DOIprefix\doi{10.1016/j.icarus.2018.12.012}.
\bibitem[{Ogilvie and Lin(2004)}]{Ogilvie2004}
\bibinfo{author}{Ogilvie, G.I.}, \bibinfo{author}{Lin, D.N.C.},
  \bibinfo{year}{2004}.
\newblock \bibinfo{title}{Tidal {{Dissipation}} in {{Rotating Giant Planets}}}.
\newblock \bibinfo{journal}{The Astrophysical Journal} \bibinfo{volume}{610},
  \bibinfo{pages}{477--509}.
\newblock \DOIprefix\doi{10.1086/421454}.
\bibitem[{Papaloizou and Larwood(2000)}]{Papaloizou2000}
\bibinfo{author}{Papaloizou, J.C.B.}, \bibinfo{author}{Larwood, J.D.},
  \bibinfo{year}{2000}.
\newblock \bibinfo{title}{On the orbital evolution and growth of protoplanets
  embedded in a gaseous disc}.
\newblock \bibinfo{journal}{Monthly Notices of the Royal Astronomical Society}
  \bibinfo{volume}{315}, \bibinfo{pages}{823--833}.
\newblock \DOIprefix\doi{10.1046/j.1365-8711.2000.03466.x}.
\bibitem[{Plescia and Boyce(1983)}]{Plescia1983}
\bibinfo{author}{Plescia, J.B.}, \bibinfo{author}{Boyce, J.M.},
  \bibinfo{year}{1983}.
\newblock \bibinfo{title}{Crater numbers and geological histories of
  {{Iapetus}}, {{Enceladus}}, {{Tethys}} and {{Hyperion}}}.
\newblock \bibinfo{journal}{Nature} \bibinfo{volume}{301},
  \bibinfo{pages}{666--670}.
\newblock \DOIprefix\doi{10.1038/301666a0}.
\bibitem[{Polycarpe et~al.(2018)Polycarpe, Saillenfest, Lainey, Vienne,
  Noyelles and Rambaux}]{Polycarpe2018}
\bibinfo{author}{Polycarpe, W.}, \bibinfo{author}{Saillenfest, M.},
  \bibinfo{author}{Lainey, V.}, \bibinfo{author}{Vienne, A.},
  \bibinfo{author}{Noyelles, B.}, \bibinfo{author}{Rambaux, N.},
  \bibinfo{year}{2018}.
\newblock \bibinfo{title}{Strong tidal energy dissipation in {{Saturn}} at
  {{Titan}}'s frequency as an explanation for {{Iapetus}} orbit}.
\newblock \bibinfo{journal}{Astronomy and Astrophysics} \bibinfo{volume}{619},
  \bibinfo{pages}{A133}.
\newblock \DOIprefix\doi{10.1051/0004-6361/201833930}.
\bibitem[{Quillen et~al.(2022)Quillen, LaBarca and Chen}]{Quillen2022}
\bibinfo{author}{Quillen, A.C.}, \bibinfo{author}{LaBarca, A.},
  \bibinfo{author}{Chen, Y.}, \bibinfo{year}{2022}.
\newblock \bibinfo{title}{Non-principal axis rotation in binary asteroid
  systems and how it weakens the {{BYORP}} effect}.
\newblock \bibinfo{journal}{Icarus} \bibinfo{volume}{374},
  \bibinfo{pages}{114826}.
\newblock \DOIprefix\doi{10.1016/j.icarus.2021.114826}.
\bibitem[{Quillen et~al.(2020)Quillen, Lane, Nakajima and Wright}]{Quillen2020}
\bibinfo{author}{Quillen, A.C.}, \bibinfo{author}{Lane, M.},
  \bibinfo{author}{Nakajima, M.}, \bibinfo{author}{Wright, E.},
  \bibinfo{year}{2020}.
\newblock \bibinfo{title}{Excitation of tumbling in {{Phobos}} and {{Deimos}}}.
\newblock \bibinfo{journal}{Icarus} \bibinfo{volume}{340},
  \bibinfo{pages}{113641}.
\newblock \DOIprefix\doi{10.1016/j.icarus.2020.113641}.
\bibitem[{Quillen et~al.(2017)Quillen, {Nichols-Fleming}, Chen and
  Noyelles}]{Quillen2017}
\bibinfo{author}{Quillen, A.C.}, \bibinfo{author}{{Nichols-Fleming}, F.},
  \bibinfo{author}{Chen, Y.Y.}, \bibinfo{author}{Noyelles, B.},
  \bibinfo{year}{2017}.
\newblock \bibinfo{title}{Obliquity evolution of the minor satellites of
  {{Pluto}} and {{Charon}}}.
\newblock \bibinfo{journal}{Icarus} \bibinfo{volume}{293},
  \bibinfo{pages}{94--113}.
\newblock \DOIprefix\doi{10.1016/j.icarus.2017.04.012}.
\bibitem[{Rein and Tamayo(2015)}]{Rein2015}
\bibinfo{author}{Rein, H.}, \bibinfo{author}{Tamayo, D.}, \bibinfo{year}{2015}.
\newblock \bibinfo{title}{{{WHFAST}}: A fast and unbiased implementation of a
  symplectic {{Wisdom-Holman}} integrator for long-term gravitational
  simulations}.
\newblock \bibinfo{journal}{Monthly Notices of the Royal Astronomical Society}
  \bibinfo{volume}{452}, \bibinfo{pages}{376--388}.
\newblock \DOIprefix\doi{10.1093/mnras/stv1257}.
\bibitem[{Seligman and Laughlin(2020)}]{Seligman2020}
\bibinfo{author}{Seligman, D.}, \bibinfo{author}{Laughlin, G.},
  \bibinfo{year}{2020}.
\newblock \bibinfo{title}{Evidence that {{1I}}/2017 {{U1}} ('{{Oumuamua}}) was
  {{Composed}} of {{Molecular Hydrogen Ice}}}.
\newblock \bibinfo{journal}{The Astrophysical Journal} \bibinfo{volume}{896},
  \bibinfo{pages}{L8}.
\newblock \DOIprefix\doi{10.3847/2041-8213/ab963f}.
\bibitem[{Shevchenko(1999)}]{Shevchenko1999}
\bibinfo{author}{Shevchenko, I.I.}, \bibinfo{year}{1999}.
\newblock \bibinfo{title}{The {{Separatrix Algorithmic Map}}: {{Application}}
  to the {{Spin-Orbit Motion}}}.
\newblock \bibinfo{journal}{Celestial Mechanics and Dynamical Astronomy}
  \bibinfo{volume}{73}, \bibinfo{pages}{259--268}.
\newblock \DOIprefix\doi{10.1023/A:1008367618329}.
\bibitem[{Tamayo et~al.(2020)Tamayo, Rein, Shi and Hernandez}]{Tamayo2020a}
\bibinfo{author}{Tamayo, D.}, \bibinfo{author}{Rein, H.}, \bibinfo{author}{Shi,
  P.}, \bibinfo{author}{Hernandez, D.M.}, \bibinfo{year}{2020}.
\newblock \bibinfo{title}{{{REBOUNDx}}: A library for adding conservative and
  dissipative forces to otherwise symplectic {{N-body}} integrations}.
\newblock \bibinfo{journal}{Monthly Notices of the Royal Astronomical Society}
  \bibinfo{volume}{491}, \bibinfo{pages}{2885--2901}.
\newblock \DOIprefix\doi{10.1093/mnras/stz2870}.
\bibitem[{Terquem(2021)}]{Terquem2021}
\bibinfo{author}{Terquem, C.}, \bibinfo{year}{2021}.
\newblock \bibinfo{title}{On a new formulation for energy transfer between
  convection and fast tides with application to giant planets and solar type
  stars}.
\newblock \bibinfo{journal}{Monthly Notices of the Royal Astronomical Society}
  \bibinfo{volume}{503}, \bibinfo{pages}{5789--5806}.
\newblock \DOIprefix\doi{10.1093/mnras/stab224}.
\bibitem[{Terquem and Papaloizou(2019)}]{Terquem2019}
\bibinfo{author}{Terquem, C.}, \bibinfo{author}{Papaloizou, J.C.B.},
  \bibinfo{year}{2019}.
\newblock \bibinfo{title}{First-order mean motion resonances in two-planet
  systems: General analysis and observed systems}.
\newblock \bibinfo{journal}{Monthly Notices of the Royal Astronomical Society}
  \bibinfo{volume}{482}, \bibinfo{pages}{530--549}.
\newblock \DOIprefix\doi{10.1093/mnras/sty2693}.
\bibitem[{Thomas et~al.(2007)Thomas, Armstrong, Asmar, Burns, Denk, Giese,
  Helfenstein, Iess, Johnson, McEwen, Nicolaisen, Porco, Rappaport, Richardson,
  Somenzi, Tortora, Turtle and Veverka}]{Thomas2007}
\bibinfo{author}{Thomas, P.C.}, \bibinfo{author}{Armstrong, J.W.},
  \bibinfo{author}{Asmar, S.W.}, \bibinfo{author}{Burns, J.A.},
  \bibinfo{author}{Denk, T.}, \bibinfo{author}{Giese, B.},
  \bibinfo{author}{Helfenstein, P.}, \bibinfo{author}{Iess, L.},
  \bibinfo{author}{Johnson, T.V.}, \bibinfo{author}{McEwen, A.},
  \bibinfo{author}{Nicolaisen, L.}, \bibinfo{author}{Porco, C.},
  \bibinfo{author}{Rappaport, N.}, \bibinfo{author}{Richardson, J.},
  \bibinfo{author}{Somenzi, L.}, \bibinfo{author}{Tortora, P.},
  \bibinfo{author}{Turtle, E.P.}, \bibinfo{author}{Veverka, J.},
  \bibinfo{year}{2007}.
\newblock \bibinfo{title}{Hyperion's sponge-like appearance}.
\newblock \bibinfo{journal}{Nature} \bibinfo{volume}{448},
  \bibinfo{pages}{50--56}.
\newblock \DOIprefix\doi{10.1038/nature05779}.
\bibitem[{Thomas et~al.(1995)Thomas, Black and Nicholson}]{Thomas1995}
\bibinfo{author}{Thomas, P.C.}, \bibinfo{author}{Black, G.J.},
  \bibinfo{author}{Nicholson, P.D.}, \bibinfo{year}{1995}.
\newblock \bibinfo{title}{Hyperion: {{Rotation}}, shape, and geology from
  {{Voyager}} images.}
\newblock \bibinfo{journal}{Icarus} \bibinfo{volume}{117},
  \bibinfo{pages}{128--148}.
\newblock \DOIprefix\doi{10.1006/icar.1995.1147}.
\bibitem[{Wisdom(1987)}]{Wisdom1987}
\bibinfo{author}{Wisdom, J.}, \bibinfo{year}{1987}.
\newblock \bibinfo{title}{Rotational {{Dynamics}} of {{Irregularly Shaped
  Natural Satellites}}}.
\newblock \bibinfo{journal}{The Astronomical Journal} \bibinfo{volume}{94},
  \bibinfo{pages}{1350}.
\newblock \DOIprefix\doi{10.1086/114573}.
\bibitem[{Wisdom et~al.(1984)Wisdom, Peale and Mignard}]{Wisdom1984}
\bibinfo{author}{Wisdom, J.}, \bibinfo{author}{Peale, S.J.},
  \bibinfo{author}{Mignard, F.}, \bibinfo{year}{1984}.
\newblock \bibinfo{title}{The chaotic rotation of {{Hyperion}}}.
\newblock \bibinfo{journal}{Icarus} \bibinfo{volume}{58},
  \bibinfo{pages}{137--152}.
\newblock \DOIprefix\doi{10.1016/0019-1035(84)90032-0}.

\end{thebibliography}

\end{document}